\documentclass[journal,10pt,twoside]{IEEEtran} 
\pdfoutput=1
\usepackage{amsmath,cite,amsfonts,amssymb,psfrag,amsthm}
\usepackage{graphicx}
\usepackage{epstopdf}
\usepackage{rotating}
\usepackage{dsfont}
\usepackage{color}
\usepackage{tikz}
\usetikzlibrary{plotmarks}
\usepackage{pgfplots}
\usetikzlibrary{calc}
\usetikzlibrary{shapes,arrows}
\usetikzlibrary{decorations.markings}
\usetikzlibrary{positioning}
\pgfplotsset{compat=1.10}
\usetikzlibrary{calc}
\usetikzlibrary{shapes,arrows}
\usetikzlibrary{decorations.markings}

\usepackage[utf8]{inputenc}
\usepackage{authblk}

\usepackage{accents}
\makeatletter
\newcommand{\thickbar}{\mathpalette\@thickbar}
\newcommand{\@thickbar}[2]{{#1\mkern1.5mu\vbox{
			\sbox\z@{$#1\mkern-1.5mu#2\mkern-1.5mu$}%
			\sbox\tw@{$#1\overline{#2}$}%
			\dimen@=\dimexpr\ht\tw@-\ht\z@-.8\p@\relax
			\hrule\@height.8\p@ 
			\vskip\dimen@
			\box\z@}\mkern1.5mu}
}
\makeatother

\DeclareFontFamily{U}{mathx}{\hyphenchar\font45}
\DeclareFontShape{U}{mathx}{m}{n}{<-> mathx10}{}
\DeclareSymbolFont{mathx}{U}{mathx}{m}{n}
\DeclareMathAccent{\widebar}{0}{mathx}{"73}

\makeatletter
\newtheorem*{rep@theorem}{\rep@title}
\newcommand{\newreptheorem}[2]{%
	\newenvironment{rep#1}[1]{%
		\def\rep@title{\Cref{##1}}%
		\begin{rep@theorem}}%
		{\end{rep@theorem}}}
\newcommand*{\textlabel}[2]{%
	\edef\@currentlabel{#1}
	\phantomsection
	#1\label{#2}
}
\makeatother

\newtheorem{theorem}{Theorem}

\newtheorem{remark}{Remark}
\newtheorem{example}{Example}
\newtheorem{definition}{Definition}

\newtheorem{lemma}{Lemma}
\newtheorem{corollary}{Corollary}

\begin{document}
\title{Multi-Entity and Multi-Enrollment Key Agreement with Correlated Noise}
\IEEEoverridecommandlockouts

\author{Onur G\"unl\"u,~\IEEEmembership{Member,~IEEE} 
\thanks{Manuscript received April 30, 2020; revised August 19, 2020 and September 15, 2020; accepted September 23, 2020. O. G\"unl\"u is supported by the German Federal Ministry of Education and Research (BMBF) within the national initiative for ``Post Shannon Communication (NewCom)'' under the Grant 16KIS1004. The associate editor coordinating the review of this manuscript and approving it for publication was Dr. Matthieu Bloch.}
\thanks{O. G\"unl\"u is with the Information Theory and Applications Chair, Technische Universit\"at Berlin, 10623 Berlin, Germany (e-mail: guenlue@tu-berlin.de).}
}

\maketitle

\begin{abstract}
	A basic model for key agreement with a remote (or hidden) source is extended to a multi-user model with joint secrecy and privacy constraints over all entities that do not trust each other after key agreement. Multiple entities using different measurements of the same source through broadcast channels (BCs) to agree on mutually-independent local secret keys are considered. Our model is the proper multi-user extension of the basic model since the encoder and decoder pairs are not assumed to trust other pairs after key agreement, unlike assumed in the literature. Strong secrecy constraints imposed on all secret keys jointly, which is more stringent than separate secrecy leakage constraints for each secret key considered in the literature, are satisfied. Inner bounds for maximum key rate, and minimum privacy-leakage and database-storage rates are proposed for any finite number of entities. Inner and outer bounds for degraded and less-noisy BCs are given to illustrate cases with strong privacy. A multi-enrollment model that is used for common physical unclonable functions is also considered to establish inner and outer bounds for key-leakage-storage regions that differ only in the Markov chains imposed. For this special case, the encoder and decoder measurement channels have the same channel transition matrix and secrecy leakage is measured for each secret key separately. We illustrate cases for which it is useful to have multiple enrollments as compared to a single enrollment and vice versa.
\end{abstract}


\IEEEpeerreviewmaketitle
\section{Introduction} \label{sec:intro}
A natural source of randomness is biometric identifiers such as fingerprints that are generally transformed into a frequency domain and quantized to obtain bit sequences that are unique to an individual \cite{Campisi}. Similarly, physical identifiers such as fine variations of ring oscillator (RO) outputs or random start-up values of static random access memories (SRAMs) that are caused by uncontrollable manufacturing variations, are safer and cheaper alternatives to key storage in a non-volatile memory \cite{GassendThesis}. Physical identifiers for digital devices such as Internet-of-Things (IoT) devices can be implemented using physical unclonable functions (PUFs) \cite{GassendThesis}. One can use PUFs in various coding schemes as a source of local randomness \cite[Chapter 1]{benimdissertation}, e.g., in the randomized encoder of the wiretap channel \cite{WTC} and of the strong coordination problem \cite{CuffStrongCoord,GiuliaStrongCoord}.

We use the basic source model for key agreement from \cite{AhlswedeCsiz,Maurer} to find achievable rate regions for key agreement with PUFs and biometric identifiers. In this classic model, an encoder observes a source output to generate a secret key and sends public side information, i.e., \textit{helper data}, to a decoder, so the decoder can reliably reconstruct the same secret key by observing another source output and the helper data. The main constraints are that the information leaked about the secret key, i.e., \emph{secrecy leakage}, is negligible and the information leaked about the identifier output, i.e., \emph{privacy leakage}, is small \cite{IgnaTrans,LaiTrans}. Furthermore, the amount of public storage should also be minimized to limit the hardware cost \cite{csiszarnarayan}.

Suppose the encoder generates a key from a noisy measurement of a hidden (or remote) source output, and a decoder has access to another noisy measurement of the same source and the helper data to reconstruct the same key. We call this model the \emph{generated-secret} (GS) model with a hidden source. This model is introduced in \cite{bizimMMMMTIFS} as an extension of the visible (noiseless) source outputs observed by the encoder, considered in \cite{IgnaTrans,LaiTrans}. Similarly, for the \emph{chosen-secret} (CS) model, an embedded (or chosen) key and noisy identifier measurements are combined by the encoder to generate the public helper data. We consider both models to address different applications.

\subsection{Related Work and Motivation}
The same identifier is used by multiple encoder and decoder pairs in \cite{LifengTransMultipleUse}, where the identifier outputs observed by different encoders are the same because the encoder measurements are assumed to be noiseless. Therefore, the multiple use of the same noiseless source output allows all encoders to know the secret key of the other encoders. This model does not fit well to the practical key agreement with identifier scenarios because there is noise in every identifier measurement.

Multiple enrollments of a hidden source using noisy measurements are considered in \cite{LienekeTIFS2019}, where weakly secure secret keys are generated without privacy leakage and storage constraints. Furthermore, there is a causality assumption in \cite{LienekeTIFS2019} on the availability of the helper data, i.e., any decoder has access to all previously-generated helper data. This assumption is not necessarily realistic as a decoder of, e.g., an IoT device that embodies a PUF should be low complexity and the amount of data to process increases linearly with the number of enrollments. In addition, any manipulation in any of the helper data can cause the complete multi-enrollment system to fail. 

A classic method used for key agreement, i.e., the fuzzy commitment scheme (FCS) \cite{FuzzyCommitment}, is used in \cite{Lieneke} in combination with an SRAM PUF to enroll the noisy outputs of the same SRAM multiple times. The symmetry condition in \cite[Eq. (16)]{Lieneke} conditioned on a fixed SRAM cell state is entirely similar to the symmetry satisfied by binary-input symmetric output (BISO) channels; see e.g., \cite[p. 613]{infcombining}, \cite[Eq. (14)]{bizimMMMMTIFS}. For SRAM outputs that satisfy this symmetry, the normalized (weak) secrecy leakage about each separate secret key is shown to be zero. It is discussed in \cite[Section 3.4]{bizimBenelux} that any uniformly-distributed hidden identifier output with BISO measurement channels satisfies the results in \cite{Lieneke}. In \cite[Theorem 1]{bizimBenelux} the secret-key capacity of the two-enrollment key agreement problem is established for measurement channels with the same channel transition matrix. However, these multi-enrollment models do not consider the privacy leakage and storage constraints, there is no constraint on the independence of the secret keys of different enrollments, and the secrecy leakage constraint is weak and is not applied jointly on all secret keys. Furthermore, optimal random linear code constructions that achieve the boundaries of the key-leakage-storage regions are given in \cite{bizimWZ}, where the classic code constructions FCS and code-offset fuzzy extractors \cite{Dodis2008fuzzy} are shown to be strictly suboptimal. Therefore, the multi-enrollment models and constructions in the literature are strictly suboptimal and not necessarily realistic. We therefore list stronger secrecy constraints jointly on all entities, which approximates the reality better in combination with storage rate and joint privacy-leakage rate constraints. These constraints define the \textit{multi-entity key agreement} problem, where the entities that use the same identifier do not have to trust other entities after key agreement. Therefore, the multi-entity key agreement problem is a proper multi-user extension of single-enrollment models. We first consider the multi-entity key agreement problem and then analyze a special case of the multi-enrollment key agreement problem to illustrate scenarios for which a single enrollment can be more useful than multiple enrollments and vice versa.

Every measurement of an identifier is considered to be noisy due to, e.g., local temperature and voltage changes in the hardware of the PUF circuit or a cut on the finger. Noise components at the encoder and decoder measurements of a hidden source can be also correlated due to, e.g., the surrounding logic in the hardware \cite{MerliROCorrelated} or constant fingertip moisture. This correlation between the noise sequences is modeled in \cite{bizimITW} as a broadcast channel (BC) \cite{CoverandThomas} with an input that is the hidden source output and with outputs that are the noisy encoder and decoder measurements. We use this model for multi-entity key agreement with identifiers, where each entity (i.e., each encoder and decoder pair) observes noisy identifier outputs of the same hidden source through different BCs. For the multi-entity key agreement problem, we allow the BCs to be different as honest entities generally use different hardware implementations of the encoder and decoder pairs, which results in different correlations between noise components. 

We also consider physically-degraded (PD) and less-noisy (LN) BCs to give finer inner and outer bounds to the key-leakage-storage regions for the GS and CS models of the multi-entity key agreement problem. For the considered PD and LN BCs, we prove that strong privacy can be achieved. In \cite{IgnaTrans,LaiTrans,MatthieuPolar}, an extra common randomness that is available to the encoder and decoder and that is hidden from the eavesdropper is required to obtain strong privacy. This assumption is not realistic since such a common randomness requires hardware protection against invasive attacks, and if such a protection is feasible, then it is not necessary to use an identifier for key agreement. 

\subsection{Models for Identifier Outputs}
We study physical and biometric identifier outputs that are independent and identically distributed (i.i.d.) according to a given probability distribution. These models are reasonable if one uses transform-coding algorithms from \cite{bizimMDPI} that occupy a small hardware area to extract almost i.i.d. bits from PUFs under varying environmental conditions. Similar transform-coding based algorithms have been applied to biometric identifiers to obtain independent output symbols \cite{Transformbio}. These transform-coding algorithms provide almost i.i.d. identifier outputs and noise sequences; however, the correlation between the noise components on the encoder and decoder components are not removed using these methods. Furthermore, PUFs are used for on-demand key reconstruction and physical attacks on PUFs permanently change the identifier outputs \cite{PappuThesis}, so we assume that the eavesdropper cannot obtain information correlated with the PUF outputs, unlike biometric identifiers.

\subsection{Summary of Contributions}
We extend the key-leakage-storage rate tuple analysis of the single-enrollment model for hidden identifier outputs measured through general BCs in \cite{bizimITW} to consider multi-entity and multi-enrollment key agreement with a set of stringent secrecy constraints. A summary of the main contributions is as follows.

\begin{itemize}
	\item We derive achievable key-leakage-storage rate tuples for the GS model with strong secrecy for any finite number of entities using the same identifier's measurements through different BCs for key agreement. Separate identifier measurements considered in \cite{bizimKittipongTIFS,bizimMMMMTIFS} correspond to a PD BC and the visible source model in \cite{IgnaTrans,LaiTrans} corresponds to a semi-deterministic BC. 
	\item For a set of PD and  LN BCs, the privacy-leakage rates for the two-entity key agreement problem are calculated. These PD and LN BCs are shown to provide strong privacy without the need of a common randomness. An outer bound is given for the considered PD and LN BCs.
	\item We next consider a special case of the multi-enrollment key agreement problem, where all measurement channels are separate (i.e., PD BCs) and they have the same transition matrix. This is a common model used for SRAM PUFs. Using a less stringent secrecy leakage constraint that bounds the information leakage for each secret key separately and without the mutual independence constraint on the secret keys, we establish inner and outer bounds for the strong-secrecy key-leakage-storage region for this two-enrollment key agreement problem. The bounds differ only in the Markov chains imposed. This result is a significant improvement to the two-enrollment secret-key rate region (without storage and privacy-leakage rate constraints) established in \cite{bizimBenelux} for weak secrecy, which is recovered by eliminating auxiliary random variables in the proposed rate regions.
	\item All inner and outer bounds for the GS model are extended to the CS model, which comprises secret-key binding methods that embed a chosen secret key to the encoder. 
	\item We give two scenarios to compare single-enrollment and two-enrollment models and illustrate that for different assumptions on measurement channels, either of the two models can perform better in terms of the privacy-leakage vs. secret-key rate boundary tuples.
\end{itemize}

\subsection{Organization}
This paper is organized as follows. In Section~\ref{sec:problem_setting}, we describe the multi-entity key agreement problem with BC measurements. We give achievable key-leakage-storage regions for the GS and CS models with strong secrecy and BC measurements for any finite number of entities in Section~\ref{sec:achievablescheme} in addition to inner and outer bounds for PD and LN BCs that satisfy strong privacy. The proposed inner bounds for the two-enrollment key agreement problem in Section~\ref{sec:tightregions} are shown to differ from the outer bounds only in the Markov chains imposed for a special case with less stringent secrecy constraints. In Sections~\ref{sec:achinnergeneral} and ~\ref{sec:Twoenrollmentproofs}, proofs of the given rate regions for the general multi-entity key agremeent problem and for the two-enrollment key agreement problem, respectively, are given. Section~\ref{sec:conclusion} concludes the paper.

\subsection{Notation}
Upper case letters represent random variables and lower case letters their realizations. A superscript denotes a string of variables, e.g., $\displaystyle X^n\!=\!X_1,X_2,\ldots, X_i,\ldots, X_n$, and a subscript $i$ denotes the position of a variable in a string. A random variable $\displaystyle X$ has probability distribution $\displaystyle P_X$. Calligraphic letters such as $\displaystyle \mathcal{X}$ denote sets, set sizes are written as $\displaystyle |\mathcal{X}|$ and their complements as $\displaystyle \mathcal{X}^c$. $[1:J]$ denotes the set $\{1,2,\ldots,J\}$ for an integer $J\geq1$ and $[1:J]\setminus\{j\}$ denotes the set $\{1,2,,\ldots,j-1,j+1,\ldots,J\}$ for any $j\in[1:J]$.  $H_b(x)=-x\log x- (1-x)\log (1-x)$ is the binary entropy function, where we take logarithms to the base $2$, and $H_b^{-1}(\cdot)$ denotes its inverse with range $[0, 0.5]$. $X\sim\text{Bern}(\alpha)$ is a binary random variable with $\Pr[X=1]=\alpha$. A binary symmetric channel (BSC) with crossover probability $p$ is denoted by BSC($p$). $Q(\cdot)$ is the $Q$-function that gives the tail probability for the standard normal distribution.

\section{Multi-Entity Key Agreement Model}\label{sec:problem_setting}
Consider hidden identifier outputs $X^n$ that are i.i.d. according to a probability distribution $P_X$. The hidden (or remote) source with outputs $X^n$ is common to all honest entities that enroll the same identifier, but they observe different noisy measurements of the same hidden source. If there are a finite number $J$ of honest entities that use the same identifer, the $j$-th encoder and decoder pair observes noisy source measurements that are outputs of a BC $P_{\widetilde{X}_jY_j|X}$, with abuse of notation, for all $j\in[1:J]$, where  $\mathcal{\widetilde{X}}_j$, $\mathcal{Y}_j$, and $\mathcal{X}$ are finite sets.

\begin{figure}
	\centering
	\resizebox{1\linewidth}{!}{
		\begin{tikzpicture}
		\node (so) at (-2.5,-4.5) [draw,rounded corners = 5pt, minimum width=0.4cm,minimum height=0.6cm, align=left] {$P_X$};
		\node (a) at (1.5,-1.0) [draw,rounded corners = 6pt, minimum width=2.10cm,minimum height=1.0cm, align=left] {$
			(S_1,W_1) \overset{(a)}{=} f_{\text{GS},1}(\widetilde{X}^n_1)$\\ $W_1\! \overset{(b)}{=}\! f_{\text{CS},1}(\widetilde{X}^n_1,S_1)$};
		\node (kdb) at (1.5,2.6) [draw,rounded corners = 6pt, minimum width=1.0cm,minimum height=1.0cm, align=left] {$\;\;$ Key\\ Database};
		\node (hddb) at (3.9,1.9) [draw,rounded corners = 6pt, minimum width=1.0cm,minimum height=1.0cm, align=left] {$\;\;$ Public\\ Database};	
		\node (comp) at (6.2,1.65) [draw,rounded corners = 6pt, minimum width=0.6cm,minimum height=0.6cm, align=left] {$=$};
		\node (quest) [right of = comp, node distance = 1.1cm] {?};		
		\node (f) at (1.5,-3.5) [draw,rounded corners = 5pt, minimum width=1.2cm,minimum height=0.9cm, align=left] {$P_{\widetilde{X}_1Y_1|X}$};
		\node (b) at (6,-1.0) [draw,rounded corners = 6pt, minimum width=3.2cm,minimum height=1.1cm, align=left] {$\hat{S}_1 = g_1\left(Y^n_1,W_1\right)$};
		\node (a1) [right of = so, node distance = 1.35cm] {$X^n$};
		\node (b1) [below of = b, node distance = 2.5cm] {$Y^n_1$};
		\node (a2) [above of = a, node distance = 1.6cm] {$S_1$};
		\node (svhat) [right of = kdb, node distance = 4.3cm] {$S_1$};		
		\node (w5) [right of = a2, node distance = 2.40cm] {$W_1$};
		\node (shat) [right of = w5, node distance = 2.0cm] {$\hat{S}_1$};				
		\draw[decoration={markings,mark=at position 1 with {\arrow[scale=1.5]{latex}}},
		postaction={decorate}, thick, shorten >=1.4pt] (so.east) -- (a1.west);
		\draw[decoration={markings,mark=at position 1 with {\arrow[scale=1.5]{latex}}},
		postaction={decorate}, thick, shorten >=1.4pt] (a1.east) -- ($(a1.east)+(0.5,0.0)$)-- ($(f.west)-(1.04,0.01)$)|- (f.west);
		\draw[decoration={markings,mark=at position 1 with {\arrow[scale=1.5]{latex}}},
		postaction={decorate}, thick, shorten >=1.4pt] (f.north) -- (a.south) node [midway, right] {$\widetilde{X}^n_1$};
		\draw[decoration={markings,mark=at position 1 with {\arrow[scale=1.5]{latex}}},
		postaction={decorate}, thick, shorten >=1.4pt] (f.east) -- (b1.west);
		\draw[decoration={markings,mark=at position 1 with {\arrow[scale=1.5]{latex}}},
		postaction={decorate}, thick, shorten >=1.4pt] (b1.north) -- (b.south);
		\draw[decoration={markings,mark=at position 1 with {\arrow[scale=1.5]{latex}}}, postaction={decorate}, thick, shorten >=1.4pt] (a.east) -- ($(b.west)-(0.87,0.01)$) -|($(w5.south)-(0.2,0)$);
		\draw[decoration={markings,mark=at position 1 with {\arrow[scale=1.5]{latex}}}, postaction={decorate}, thick, shorten >=1.4pt] ($(w5.south)+(0.18,0)$) -- ($(b.west)-(0.32,0.01)$)|- (b.west);
		\draw[decoration={markings,mark=at position 1 with {\arrow[scale=1.5]{latex}}},
		postaction={decorate}, thick, shorten >=1.4pt] ($(w5.north)-(0.2,0)$) -- ($(hddb.south)-(0.2,0)$);
		\draw[decoration={markings,mark=at position 1 with {\arrow[scale=1.5]{latex}}},
		postaction={decorate}, thick, shorten >=1.4pt] ($(hddb.south)+(0.2,0)$) -- ($(w5.north)+(0.2,0)$);
		\draw[decoration={markings,mark=at position 1 with {\arrow[scale=1.5]{latex}}},
		postaction={decorate}, thick, shorten >=1.4pt]  ($(a2.south)+(0.2,0)$)-- ($(a.north)+(0.2,0)$) node [midway, right] {$(b)$};
		\draw[decoration={markings,mark=at position 1 with {\arrow[scale=1.5]{latex}}},
		postaction={decorate}, thick, shorten >=1.4pt] ($(a.north)-(0.2,0)$)-- ($(a2.south)-(0.2,0)$) node [midway, left] {$(a)$};
		\draw[decoration={markings,mark=at position 1 with {\arrow[scale=1.5]{latex}}},
		postaction={decorate}, thick, shorten >=1.4pt] ($(a2.north)-(0.2,0)$)-- ($(kdb.south)-(0.2,0)$) node [midway, left] {$(a)$};
		\draw[decoration={markings,mark=at position 1 with {\arrow[scale=1.5]{latex}}},
		postaction={decorate}, thick, shorten >=1.4pt]  ($(kdb.south)+(0.2,0)$)-- ($(a2.north)+(0.2,0)$) node [midway, right] {$(b)$};
		\draw[decoration={markings,mark=at position 1 with {\arrow[scale=1.5]{latex}}} ,
		postaction={decorate}, thick, shorten >=1.4pt]  ($(kdb.east)+(0,0)$)-- ($(svhat.west)+(0,0)$);
		\draw[decoration={markings,mark=at position 1 with {\arrow[scale=1.5]{latex}}},
		postaction={decorate}, thick, shorten >=1.4pt]  ($(b.north)-(0.1,0)$)-- ($(shat.south)+(0,0)$);
		\draw[decoration={markings,mark=at position 1 with {\arrow[scale=1.5]{latex}}},
		postaction={decorate}, thick, shorten >=1.4pt]  ($(shat.north)+(0,0)$)-- (comp.south);
		\draw[decoration={markings,mark=at position 1 with {\arrow[scale=1.5]{latex}}},
		postaction={decorate}, thick, shorten >=1.4pt]  ($(svhat.south)-(0,0)$)-- (comp.north);
		\draw[decoration={markings,mark=at position 1 with {\arrow[scale=1.5]{latex}}},
		postaction={decorate}, thick, shorten >=1.4pt]  (comp.east) -- (quest.west);
		\node (a5) at (1.5,-8) [draw,rounded corners = 6pt, minimum width=2.10cm,minimum height=1.0cm, align=left] {$
			(S_2,W_2) \overset{(a)}{=} f_{\text{GS},2}(\widetilde{X}^n_2)$\\ $W_2\! \overset{(b)}{=}\! f_{\text{CS},2}(\widetilde{X}^n_2,S_2)$};
		\node (kdb5) at (1.5,-11.6) [draw,rounded corners = 6pt, minimum width=1.0cm,minimum height=1.0cm, align=left] {$\;\;$ Key\\ Database};
		\node (hddb5) at (3.9,-10.9) [draw,rounded corners = 6pt, minimum width=1.0cm,minimum height=1.0cm, align=left] {$\;\;$ Public\\ Database};	
		\node (comp5) at (6.2,-10.65) [draw,rounded corners = 6pt, minimum width=0.6cm,minimum height=0.6cm, align=left] {$=$};
		\node (quest5) [right of = comp5, node distance = 1.1cm] {?};		
		\node (f5) at (1.5,-5.5) [draw,rounded corners = 5pt, minimum width=1.2cm,minimum height=0.9cm, align=left] {$P_{\widetilde{X}_2Y_2|X}$};
		\node (b5) at (6,-8) [draw,rounded corners = 6pt, minimum width=3.2cm,minimum height=1.1cm, align=left] {$\hat{S}_2 = g_2\left(Y^n_2,W_2\right)$};
		\node (b15) [above of = b5, node distance = 2.5cm] {$Y^n_2$};
		\node (a25) [below of = a5, node distance = 1.6cm] {$S_2$};
		\node (svhat5) [right of = kdb5, node distance = 4.3cm] {$S_2$};		
		\node (w55) [right of = a25, node distance = 2.40cm] {$W_2$};
		\node (shat5) [right of = w55, node distance = 2.0cm] {$\hat{S}_2$};				
		\draw[decoration={markings,mark=at position 1 with {\arrow[scale=1.5]{latex}}},
		postaction={decorate}, thick, shorten >=1.4pt] (a1.east) -- ($(a1.east)+(0.5,0.0)$)-- ($(f5.west)-(1.04,0.01)$)|- (f5.west);
		\draw[decoration={markings,mark=at position 1 with {\arrow[scale=1.5]{latex}}},
		postaction={decorate}, thick, shorten >=1.4pt] (f5.south) -- (a5.north) node [midway, right] {$\widetilde{X}^n_2$};
		\draw[decoration={markings,mark=at position 1 with {\arrow[scale=1.5]{latex}}},
		postaction={decorate}, thick, shorten >=1.4pt] (f5.east) -- (b15.west);
		\draw[decoration={markings,mark=at position 1 with {\arrow[scale=1.5]{latex}}},
		postaction={decorate}, thick, shorten >=1.4pt] (b15.south) -- (b5.north);
		\draw[decoration={markings,mark=at position 1 with {\arrow[scale=1.5]{latex}}}, postaction={decorate}, thick, shorten >=1.4pt] (a5.east) -- ($(b5.west)-(0.87,0.01)$) -|($(w55.north)-(0.2,0)$);
		\draw[decoration={markings,mark=at position 1 with {\arrow[scale=1.5]{latex}}}, postaction={decorate}, thick, shorten >=1.4pt] ($(w55.north)+(0.18,0)$) -- ($(b5.west)-(0.32,0.01)$)|- (b5.west);
		\draw[decoration={markings,mark=at position 1 with {\arrow[scale=1.5]{latex}}},
		postaction={decorate}, thick, shorten >=1.4pt] ($(w55.south)-(0.2,0)$) -- ($(hddb5.north)-(0.2,0)$);
		\draw[decoration={markings,mark=at position 1 with {\arrow[scale=1.5]{latex}}},
		postaction={decorate}, thick, shorten >=1.4pt] ($(hddb5.north)+(0.2,0)$) -- ($(w55.south)+(0.2,0)$);
		\draw[decoration={markings,mark=at position 1 with {\arrow[scale=1.5]{latex}}},
		postaction={decorate}, thick, shorten >=1.4pt]  ($(a25.north)+(0.2,0)$)-- ($(a5.south)+(0.2,0)$) node [midway, right] {$(b)$};
		\draw[decoration={markings,mark=at position 1 with {\arrow[scale=1.5]{latex}}},
		postaction={decorate}, thick, shorten >=1.4pt] ($(a5.south)-(0.2,0)$)-- ($(a25.north)-(0.2,0)$) node [midway, left] {$(a)$};
		\draw[decoration={markings,mark=at position 1 with {\arrow[scale=1.5]{latex}}},
		postaction={decorate}, thick, shorten >=1.4pt] ($(a25.south)-(0.2,0)$)-- ($(kdb5.north)-(0.2,0)$) node [midway, left] {$(a)$};
		\draw[decoration={markings,mark=at position 1 with {\arrow[scale=1.5]{latex}}},
		postaction={decorate}, thick, shorten >=1.4pt]  ($(kdb5.north)+(0.2,0)$)-- ($(a25.south)+(0.2,0)$) node [midway, right] {$(b)$};
		\draw[decoration={markings,mark=at position 1 with {\arrow[scale=1.5]{latex}}} ,
		postaction={decorate}, thick, shorten >=1.4pt]  ($(kdb5.east)+(0,0)$)-- ($(svhat5.west)+(0,0)$);
		\draw[decoration={markings,mark=at position 1 with {\arrow[scale=1.5]{latex}}},
		postaction={decorate}, thick, shorten >=1.4pt]  ($(b5.south)-(0.1,0)$)-- ($(shat5.north)+(0,0)$);
		\draw[decoration={markings,mark=at position 1 with {\arrow[scale=1.5]{latex}}},
		postaction={decorate}, thick, shorten >=1.4pt]  ($(shat5.south)+(0,0)$)-- (comp5.north);
		\draw[decoration={markings,mark=at position 1 with {\arrow[scale=1.5]{latex}}},
		postaction={decorate}, thick, shorten >=1.4pt]  ($(svhat5.north)-(0,0)$)-- (comp5.south);
		\draw[decoration={markings,mark=at position 1 with {\arrow[scale=1.5]{latex}}},
		postaction={decorate}, thick, shorten >=1.4pt]  (comp5.east) -- (quest5.west);
		\end{tikzpicture}
	}
	\caption{Illustration of the multi-entity key agremeent problem for $J=2$ entities with encoder and decoder measurements through BCs for $(a)$ the GS model and $(b)$ the CS model.}\label{fig:ProblemDefinitionfortwo}
\end{figure}
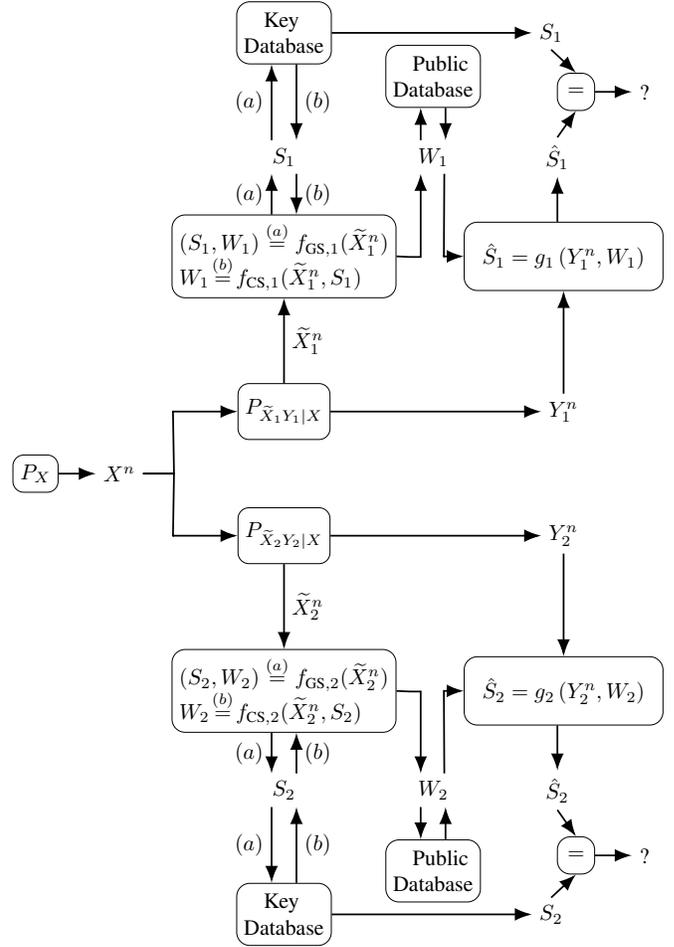

For the GS model illustrated in Fig.~\ref{fig:ProblemDefinitionfortwo}$(a)$ for $J=2$ honest entities, the $j$-th encoder $f_{\text{GS},j}(\cdot)$ generates helper data $W_j$ and a secret key $S_j$ from its observed sequence $\widetilde{X}^n_j$. All secret keys are stored in a secure database, whereas helper data are stored in a public database so that an eavesdropper has access only to the helper data. Using the helper data $W_j$ and its observed sequence $Y^n_j$, the $j$-th decoder $g_j(\cdot,\cdot)$ generates the key estimate $\hat{S}_j$. Similar steps are applied for the CS model in Fig.~\ref{fig:ProblemDefinitionfortwo}$(b)$ also for $J=2$ honest entities, except that each $S_j$ should be embedded into the $j$-th encoder $f_{\text{CS},j}(\cdot,\cdot)$.

Denote a set of secret keys as
\begin{align}
\mathcal{S}_{\mathcal{K}} = \{S_j:j\in\mathcal{K}\} 
\end{align}
and a set of helper data as
\begin{align}
\mathcal{W}_{\mathcal{K}} = \{W_j:j\in\mathcal{K}\}
\end{align}
for any $\mathcal{K} \subseteq[1:J]$. A (secret-key, privacy-leakage, storage), or key-leakage-storage, rate tuple is denoted as $(R_{\text{s}}, R_{\ell},R_{\text{w}})$. Similarly, we denote a set of secret-key rates, for any $\mathcal{K} \subseteq[1:J]$, as
\begin{align}
\mathcal{R}_{\text{s},\mathcal{K}} = \{R_{\text{s},j}:j\in\mathcal{K}\}
\end{align}
and a set of storage rates as
\begin{align}
\mathcal{R}_{\text{w},\mathcal{K}} = \{R_{\text{w},j}:j\in\mathcal{K}\}.
\end{align}

We next define the multi-entity key-leakage-storage regions.

\begin{definition}
\normalfont A key-leakage-storage rate tuple $(\mathcal{R}_{\text{s},[1:J]} , R_{\ell},\mathcal{R}_{\text{w},[1:J]} )$ is achievable for the multi-entity GS and CS models with $j$-th encoder and decoder measurements through a BC $P_{\widetilde{X}_jY_j|X}$ if, given any $\delta\!>\!0$, there is some $n\!\geq\!1$, and $J$ encoder and decoder pairs for which $\displaystyle R_{\text{s},j}=\frac{\log|\mathcal{S}_j|}{n}$ for all $ j\in[1:J]$ and
\begin{alignat}{2}
		&\Pr\left[\underset{j\in[1:J]}{\bigcup}\{S_j\ne\hat{S}_j\}\right] \leq \delta &&\quad \text{(reliability)}\label{eq:reliabilityconst}\\
		&\frac{1}{n}H(S_j) \geq R_{\text{s},j}-\delta,\quad\;\;\, \forall j\!\in\![1\!:\!J]&&\quad \text{(key uniformity)}\label{eq:uniformityconst}\\[2pt]
		&I\left(\mathcal{S}_{\mathcal{K}};\mathcal{S}_{\mathcal{K}^c}\right)\leq \delta,\qquad\;\;\;\;\; \forall \mathcal{K}\!\subseteq\![1\!:\!J]&&\quad \text{(strong key ind.)}\label{eq:keyindependence}\\[2pt]
		&\frac{1}{n}I(X^n;\mathcal{W}_{[1:J]})\! \leq\! R_{\ell}\!+\!\delta  &&\quad \text{(privacy)} \label{eq:privacyconst}\\[2pt]
		&I\left(\mathcal{S}_{[1:J]};\mathcal{W}_{[1:J]}\right) \leq \delta &&\quad\text{(strong secrecy)} \label{eq:secrecyconst}\\[2pt]
		&\frac{1}{n} \log|\mathcal{W}_j| \leq R_{\text{w},j}\!+\!\delta,\quad \forall j\!\in\![1\!:\!J] &&\quad \text{(storage)}\label{eq:storageconst}.
\end{alignat}

The \emph{multi-entity key-leakage-storage} regions $\mathcal{C}_{\text{gs}}$ for the GS model and $\mathcal{C}_{\text{cs}}$ for the CS model are the closures of the set of all achievable rate tuples $(\mathcal{R}_{\text{s},[1:J]} , R_{\ell},\mathcal{R}_{\text{w},[1:J]})$.

\end{definition}

Both secret-key uniformity (\ref{eq:uniformityconst}) and storage rate (\ref{eq:storageconst}) constraints correspond to $J$ separate constraints. However, reliability (\ref{eq:reliabilityconst}), strong and mutual key independence (\ref{eq:keyindependence}), privacy-leakage rate (\ref{eq:privacyconst}), and secrecy leakage (\ref{eq:secrecyconst}) constraints are joint constraints for all $J$ honest entities. Suppose after a key generation, an honest entity has access only to its corresponding secret key and it does not have access to other entities’ keys or sequences or even to the sequence it observed to generate its secret key. 

The mutual key independence constraint in (\ref{eq:keyindependence}) is not imposed in the multi-enrollment key agreement problem considered in \cite{Lieneke}. Furthermore, a normalized (weak) version of this constraint is imposed in the multi-enrollment key agreement problem considered in \cite{LienekeTIFS2019}, where the $j$-th decoder $g_j(\cdot,\cdot)$ is assumed to have access to the set of helper data $\mathcal{W}_{[1:j]}$ for all $j\in[1:J]$. The lack of the mutual key independence constraint and the assumption of availability of all previous helper data require that different encoder and decoder pairs should trust each other after key agreement. This can be the case, e.g., if all enrollments are made by the same entity. Therefore, the multi-entity key agreement problem imposes strictly more stringent constraints than the multi-enrollment key agreement problem.

The unnormalized secrecy leakage constraint (\ref{eq:secrecyconst}) provides strong secrecy, which is a stronger notion than the weak secrecy considered in \cite{IgnaTrans, LaiTrans, bizimMMMMTIFS,bizimKittipongTIFS,Lieneke,LienekeTIFS2019}. Furthermore, (\ref{eq:secrecyconst}) is more stringent than the set of individual secrecy leakage constraints $I(S_j;\mathcal{W}_{[1:J]})$ imposed for all $j\in[1:J]$, considered in \cite{Lieneke} for symmetric SRAM PUF outputs in combination with the suboptimal FCS. 

The unnormalized privacy leakage $I(X^n;\mathcal{W}_{[1:J]})$ cannot be bounded by a finite number in general. We illustrate special strong privacy cases in the next section.

\section{Inner Bounds}\label{sec:achievablescheme}
We are interested in characterizing the optimal trade-off among the secret-key, privacy-leakage, and storage rates with strong secrecy for BC measurements at the encoders and decoders of any finite number $J$ of entities that use the same hidden identifier outputs for the multi-entity key agreement problem. We give achievable rate regions for the GS and CS models in Theorem~\ref{theo:GeneralInnergscs}. The proofs are given in Section~\ref{sec:achinnergeneral}. 

Denote
\begin{align}
\mathcal{U}_{\mathcal{K}} = \{U_j:j\in\mathcal{K}\}
\end{align}
and define a function $\max\{\cdot,\cdot\}$ that gives the maximum of the input values as its output.

\begin{theorem}[Inner Bounds for Multi-entity GS and CS Models]\label{theo:GeneralInnergscs}
	An achievable rate region $\mathcal{R}_{\text{gs}}$ for the multi-entity GS model with $J$ entities is the union over all $P_{U_{j}|\widetilde{X}_j}$ for all $j\in [1:J]$ of the rate tuples such that $R_{\text{s},j} \geq 0$ for all $j \in [1:J]$ and
	\begin{alignat}{2}
	&R_{\text{s},j}\leq I(U_j;Y_j)-I(U_j;U_{[1:J]\setminus\{j\}}), &&\;\; \forall j\in [1:J]\label{eq:corrkeyrate}\\
	&R_\ell \geq \sum_{j=1}^{J}\max\{0, I(U_j;X)\!-\!I(U_j;Y_j)\}, \label{eq:corrleakagerate}\\
	&R_{\text{w},j}\! \geq\! I(U_j;\widetilde{X}_j)-I(U_j;Y_j),&& \;\;\forall j \in [1:J] \label{eq:corrstoragerate}\\
	&R_{\text{s},j}+R_{\text{w},j}\leq H(U_j|\,\mathcal{U}_{[1:J]\setminus\{j\}}),&& \;\;\forall j \in [1:J]. \label{eq:corrstorageandkeysumrate}
	\end{alignat}
	
	An achievable rate region $\mathcal{R}_{\text{cs}}$ for the multi-entity CS model with $J$ entities is the union over all $P_{U_{j}|\widetilde{X}_j}$ for all $j\in [1:J]$ of the rate tuples such that $R_{\text{s},j} \geq 0$ for all $j \in [1:J]$, (\ref{eq:corrkeyrate}), (\ref{eq:corrleakagerate}), and
	\begin{alignat}{2}
&R_{\text{w},j}\! \geq\! I(U_j;\widetilde{X}_j)-I(U_j;U_{[1:J]\setminus\{j\}}),&&\quad \forall j \in [1:J] \label{eq:cscorrstoragerate}\\
	&R_{\text{w},j}\leq H(U_j|\,\mathcal{U}_{[1:J]\setminus\{j\}}),&& \quad\forall j \in [1:J] \label{eq:cscorrstorageandkeysumrate}.
\end{alignat}
For the achievable rate regions $\mathcal{R}_{\text{gs}}$ and $\mathcal{R}_{\text{cs}}$, we have
   \begin{align}
   \displaystyle P_{\mathcal{U}_{[1:J]}\mathcal{\widetilde{X}}_{[1:J]}X\mathcal{Y}_{[1:J]}}=P_{X}\prod_{j=1}^JP_{U_j|\widetilde{X}_j}P_{\widetilde{X}_jY_j|X} \label{eq:fixedprobdistr}.
   \end{align}	
\end{theorem}

\begin{corollary}\label{cor:strongprivacy}
	Suppose for all $j\in[1:J]$ that
	\begin{itemize}
		\item $\widetilde{X}_j-Y_j-X$ form a Markov chain, i.e., $X$ is a PD version of $Y_j$ with respect to $\widetilde{X}_j$, or
		\item $P_{XY_j|\widetilde{X}_j}$ is a LN BC with $I(U_j;Y_j)\geq I(U_j;X)$ for all $P_{U_j|\widetilde{X}_j}$. 
	\end{itemize}
	For these cases, strong privacy, i.e.,
	\begin{align}
		R_\ell\geq 0 \label{eq:privacyleakzero}
	\end{align}
	can be achieved for the multi-entity GS and CS models in combination with the other corresponding bounds given in Theorem~\ref{theo:GeneralInnergscs}. 
\end{corollary}

The proof of Corollary~\ref{cor:strongprivacy} follows from Theorem~\ref{theo:GeneralInnergscs} because $I(U_j;X)-I(U_j;Y_j)\leq 0$ for all $j\in[1:J]$ for BCs considered in Corollary~\ref{cor:strongprivacy}. 

Corollary~\ref{cor:strongprivacy} illustrates that it is possible to obtain strong privacy, i.e., negligible unnormalized privacy leakage, without the requirement of a common randommness that is hidden from an eavesdropper assumed in \cite{IgnaTrans,LaiTrans,MatthieuPolar}. This is the case because the observation $Y^n_j$ of each decoder is ``better" than the observation $\widetilde{X}^n_j$ of the corresponding encoder with respect to the hidden source $X^n$ for all entities. 

\begin{remark}
	\normalfont The rate regions for our problem depend on the joint conditional probability distributions $P_{XY_j|\widetilde{X}_j}$ rather than only the marginal conditional distributions. Thus, the key-leakage-storage regions for the stochastically-degraded BCs are not necessarily equal to the regions for the corresponding PD BCs, unlike in the classic BC problem. Furthermore, since $P_{\mathcal{\widetilde{X}}_{[1:J]}X\mathcal{Y}_{[1:J]}}$ is fixed, the distinction between the LN BCs and essentially-less noisy BCs \cite{NairEssentially}, is not necessary.
\end{remark}

We next give simple outer bounds for the multi-entity key-leakage-storage regions $\mathcal{C}_{\text{gs}}$ for the GS model and $\mathcal{C}_{\text{cs}}$ for the CS model when the BCs $P_{XY_j|\widetilde{X}_j}$ for all $j\in[1:J]$ are PD BCs or LN BCs, as defined in Corollary~\ref{cor:strongprivacy}. These simple outer bounds give insights into the reason for different bounds on the secret-key rates. Based on these insights, we show a special multi-enrollment case in the next section with a less stringent secrecy constraint, for which the inner and outer bounds differ only in the Markov chains imposed and we illustrate that they match for simpler models.

\begin{lemma}\label{lem:zeroprivacyouterbound}
	Suppose one of the cases given in Corollary~\ref{cor:strongprivacy} is satisfied by the BCs $P_{XY_j|\widetilde{X}_j}$ for all $j\in[1:J]$. An outer bound on the multi-entity key-leakage-storage region $\mathcal{C}_{\text{gs}}$ is the union over all $P_{U_{j}|\widetilde{X}_{j}}$, where $U_j-\widetilde{X}_{j}-(X,Y_j)$ form a Markov chain, for all $j\in[1:J]$ of the rate tuples such that $R_{\text{s},j} \geq 0$ for all $j \in [1:J]$, (\ref{eq:corrstoragerate}), (\ref{eq:privacyleakzero}), and
	\begin{align}
	&R_{\text{s},j} \leq I(U_j;Y_j),\qquad\qquad\forall j\in[1:J]\label{eq:OuterSKRatetwoentityrall}.
	\end{align}
	
	An outer bound to the multi-entity key-leakage-storage region $\mathcal{C}_{\text{cs}}$ for the same BCs $P_{XY_j|\widetilde{X}_j}$ is the union over all $P_{U_{j}|\widetilde{X}_{j}}$, where $U_j-\widetilde{X}_{j}-(X,Y_j)$ form a Markov chain, for all $j\in[1:J]$ of the rate tuples such that $R_{\text{s},j} \geq 0$ for all $j \in [1:J]$, (\ref{eq:privacyleakzero}), (\ref{eq:OuterSKRatetwoentityrall}), and
	\begin{alignat}{2}
	&R_{\text{w},j}\geq I(U_j;\widetilde{X}_j),\qquad\qquad\forall j\in[1:J]\label{eq:Outercsstortwoentityrall}.
	\end{alignat}
\end{lemma}

The proof of Lemma~\ref{lem:zeroprivacyouterbound} follows straightforwadly by following the steps in \cite[Section VI]{bizimMMMMTIFS}, defining the auxiliary random variables $U_{j,i}= (S_j,W_j,Y_j^{i-1})$ for all $j\in[1:J]$ and $i\in[1:n]$, and by bounding $I(X^n;\mathcal{W}_{[1:J]})\geq 0$; therefore, we omit the proof.

The outer bounds do not include the inequalities in (\ref{eq:corrstorageandkeysumrate}) and (\ref{eq:cscorrstorageandkeysumrate}). Furthermore, the secret-key rate achieved by the inner bound in (\ref{eq:corrkeyrate}) is smaller than the outer bound given in (\ref{eq:OuterSKRatetwoentityrall}), where the difference is the term $-I(U_j;\mathcal{U}_{[1:J]\setminus\{j\}})$. This term is a result of the constraint in (\ref{eq:independenceofindices}) that is imposed to satisfy the strong and mutual key independence constraint given in (\ref{eq:keyindependence}). Therefore, we next consider a model without the  constraint in (\ref{eq:keyindependence}) and use a secrecy-leakage constraint that is less stringent than the one in (\ref{eq:secrecyconst}), i.e., replace (\ref{eq:secrecyconst}) by \begin{align}
I(S_j;\mathcal{W}_{[1:J]})\leq\delta,\qquad\qquad \forall j\in[1:J]
\end{align}
which is also a strong secrecy metric. Due to the lack of a mutual key independence constraint, the model in the next section is not a multi-entity model but rather a multi-enrollment model. For a special case of this multi-enrollment key agreement problem, we establish inner and outer bounds for the key-leakage-storage regions that comprise the same bounds but for different Markov chains.

\section{Bounds for a Multi-Enrollment Model}\label{sec:tightregions}
Consider next the multi-enrollment model, where the strong and mutual key independence constraint (\ref{eq:keyindependence}) of the multi-entity model is not imposed. Assume further $J=2$ entities that measure noisy outputs of the same hidden source $X^n$ through separate channels that have the same channel transition matrices, i.e., for all $j\in[1:2]$, $\tilde{x}_j\in\mathcal{\widetilde{X}}$, and $y_j\in\mathcal{\widetilde{X}}$ we have
\begin{align}
P_{\widetilde{X}_jY_j|X}(\tilde{x}_j,y_j|x)=P_{\widetilde{X}|X}(\tilde{x}_j|x)P_{\widetilde{X}|X}(y_j|x). \label{eq:SRAMchanneldef}
\end{align}
This model is common for SRAM PUFs, for which each measurement channel is modeled as a BSC with the same crossover probability corresponding to a worst case scenario\cite{RoelSRAMPUF}. Using (\ref{eq:SRAMchanneldef}), we define a multi-enrollment model.

\begin{definition}\label{def:twoenrollment}
	\normalfont A key-leakage-storage rate tuple $(\widebar{R}_{\text{s},1}, \widebar{R}_{\text{s},2}, \widebar{R}_{\ell}, \widebar{R}_{\text{w},1}, \widebar{R}_{\text{w},2})$ is achievable for the multi-enrollment GS and CS models with measurements through a BC $\displaystyle P_{\widetilde{X}Y|X}(\tilde{x},y|x)$ as in (\ref{eq:SRAMchanneldef}) if, given any $\delta\!>\!0$, there is some $n\!\geq\!1$, and two encoder and decoder pairs for which $\displaystyle \widebar{R}_{\text{s},1}=\frac{\log|\mathcal{S}_1|}{n}$, $\displaystyle \widebar{R}_{\text{s},2}=\frac{\log|\mathcal{S}_2|}{n}$, $\displaystyle\widebar{R}_{\text{w},1}=\frac{H(W_1)}{n}$, $\displaystyle\widebar{R}_{\text{w},2}=\frac{H(W_2)}{n}$, and
	\begin{alignat}{2}
	&\Pr\left[\{S_1\ne\hat{S}_1\}\bigcup\{S_2\ne\hat{S}_2\}\right] \leq \delta &&\quad \text{(reliability)}\label{eq:reliabilityconstJ2}\\[2pt]
	&\frac{1}{n}H(S_j) = \widebar{R}_{\text{s},j}-\delta,\qquad\;\;\,  j=1,2&&\quad \text{(key uniformity)}\label{eq:uniformityconstJ2}\\[2pt]
	&\frac{1}{n}I(X^n;W_1,W_2)\! =\! \widebar{R}_{\ell}\!+\!\delta  &&\quad \text{(privacy)} \label{eq:privacyconstJ2}\\[2pt]
	&I\left(S_j;W_1,W_2\right) \leq \delta, \qquad\quad j=1,2&&\quad\text{(strong secrecy)} \label{eq:secrecyconstJ2}\\[2pt]
	&\frac{1}{n} \log|\mathcal{W}_j|= \widebar{R}_{\text{w},j}\!+\!\delta,\qquad  j=1,2 &&\quad \text{(storage)}\label{eq:storageconstJ2}\\
	&I(W_1;W_2)\leq \delta&&\quad \text{(storage ind.)}.\label{eq:storageindependence}
	\end{alignat}
	
	The \emph{multi-enrollment key-leakage-storage} regions $\mathcal{\widebar{C}}_{\text{gs},J=2}$ for the GS model and $\mathcal{\widebar{C}}_{\text{cs},J=2}$ for the CS model are the closures of the set of all achievable rate tuples $(\widebar{R}_{\text{s},1} , \widebar{R}_{\text{s},2} ,\widebar{R}_{\ell},\widebar{R}_{\text{w},1},\widebar{R}_{\text{w},2})$.
	
\end{definition}

We characterize in Theorem~\ref{theo:Twoenrollmentgscs} inner and outer bounds for $\mathcal{\widebar{C}}_{\text{gs},J=2}$ and $\mathcal{\widebar{C}}_{\text{cs},J=2}$. The proofs of Theorem~\ref{theo:Twoenrollmentgscs} are given in Section~\ref{sec:Twoenrollmentproofs}, where the reason for the necessity of the secrecy-leakage constraint in (\ref{eq:secrecyconstJ2}) that is less stringent than the joint secrecy-leakage constraint in (\ref{eq:secrecyconst}) is given in Remark~\ref{rem:whylessstringent}. Similarly, the reason for the necessity of the strong helper data (storage) independence constraint in (\ref{eq:storageindependence}) is discussed in Remark~\ref{rem:helperdataindependence}. We remark that the equalities in (\ref{eq:uniformityconstJ2}), (\ref{eq:privacyconstJ2}), and (\ref{eq:storageconstJ2}) are required in the outer bounds in Theorem~\ref{theo:Twoenrollmentgscs} to provide both upper and lower bounds on $\widebar{R}_{\ell}$ and $\widebar{R}_{\text{w},j}$ in terms of Shannon entropy terms.

Denote
\begin{align}
j'= 3-j,\qquad\qquad\qquad j=1,2.\label{jprimedef}
\end{align}
 
\begin{theorem}\label{theo:Twoenrollmentgscs}
	{\normalfont (Inner Bounds for Multi-enrollment GS and CS Models)}:
	An achievable multi-enrollment key-leakage-storage region $\mathcal{\widebar{R}}_{\text{gs},J=2}$ is the union over all $P_{U_{1}|\widetilde{X}_1}$ and $P_{U_{2}|\widetilde{X}_2}$ of the rate tuples such that $\widebar{R}_{\text{s},j} \geq 0$ for $j=1,2$ and
	\begin{alignat}{2}
	&\widebar{R}_{\text{s},j}\leq I(U_j;Y_j), &&\quad j=1,2\label{eq:corrkeyrateTwoenroll}\\
	&\widebar{R}_\ell \geq \sum_{j=1}^2\big(I(U_j;X)\!-\!I(U_j;Y_j)\big), \label{eq:corrleakagerateTwoenroll}\\
	&\widebar{R}_\ell \leq \sum_{j=1}^2\big(I(U_j;X)\!-\!I(U_j;\widetilde{X}_j)\!+\!\widebar{R}_{\text{w},j}\big), \label{eq:corrleakagerateTwoenrollupperbound}\\	
	&\widebar{R}_{\text{w},j}\! \geq\! I(U_j;\widetilde{X}_j)-I(U_j;Y_j),&& \quad j=1,2 \label{eq:corrstoragerateTwoenroll}\\
	&\widebar{R}_{\text{s},j}+\widebar{R}_{\text{w},j}\leq H(U_j),&&\quad j=1,2\label{eq:sumonsamesandwpart1Twoenroll}\\
	&\widebar{R}_{\text{s},j}+\widebar{R}_{\text{w},j}+\widebar{R}_{\text{w},j'}\leq H(U_{j},U_{j'}),&&\quad j=1,2\label{eq:sumonsamesjandwjandjprimeTwoenroll}.
	\end{alignat}
	An achievable multi-enrollment key-leakage-storage region $\mathcal{\widebar{R}}_{\text{cs},J=2}$ is the union over all $P_{U_{1}|\widetilde{X}_1}$ and $P_{U_{2}|\widetilde{X}_2}$ of the rate tuples such that $\widebar{R}_{\text{s},j} \geq 0$ for $j=1,2$, (\ref{eq:corrkeyrateTwoenroll})-(\ref{eq:corrleakagerateTwoenrollupperbound}), and
	\begin{alignat}{2}
	&\widebar{R}_{\text{w},j}\! \geq\! I(U_j;\widetilde{X}_j),&&\quad j=1,2 \label{eq:cscorrstoragerateTwoenroll}\\
	&\widebar{R}_{\text{w},j}\leq H(U_j),&&\quad j=1,2\label{eq:cssumonsamesandwpart1Twoenroll}\\
	&\widebar{R}_{\text{w},j}\!+\!\widebar{R}_{\text{w},j'}\!\leq\! H(U_{j},U_{j'})\!+\!\widebar{R}_{\text{s},j'},&&\quad j=1,2\label{eq:cssumonsamesjandwjandjprimeTwoenroll}.
	\end{alignat}
	For both achievable rate regions $\mathcal{\widebar{R}}_{\text{gs},J=2}$ and $\mathcal{\widebar{R}}_{\text{cs},J=2}$, we have
	\begin{align}
	\displaystyle &P_{U_{1}U_2\widetilde{X}_{1}\widetilde{X}_2XY_{1}Y_2}({u_{1},u_2,\widetilde{x}_{1},\widetilde{x}_2,x,y_{1},y_2})\nonumber\\
	&\quad=P_{U_1|\widetilde{X}_1}(u_1|\tilde{x}_1)P_{U_2|\widetilde{X}_2}(u_2|\tilde{x}_2)P_{\widetilde{X}|X}(\tilde{x}_1|x)P_{\widetilde{X}|X}(\tilde{x}_2|x)\nonumber\\
	&\quad\qquad \times P_{\widetilde{X}|X}(y_1|x)P_{\widetilde{X}|X}(y_2|x)P_{X}(x) \label{eq:fixedprobdistrTwoenroll}.
	\end{align}	
	
	{\normalfont (Outer Bounds for Multi-enrollment GS and CS Models)} An outer bound for $\mathcal{\widebar{C}}_{\text{gs},J=2}$ is the union over all $P_{U_{1}|\widetilde{X}_{1}}$ and $P_{U_{2}|\widetilde{X}_2}$ of the rate tuples such that $\widebar{R}_{\text{s},j} \geq 0$, (\ref{eq:corrkeyrateTwoenroll}) - (\ref{eq:sumonsamesjandwjandjprimeTwoenroll}), and $U_j-\widetilde{X}_{j}-X-Y_j$ form a Markov chain for $j=1,2$. An outer bound for $\mathcal{\widebar{C}}_{\text{cs},J=2}$ is the union over all $P_{U_{1}|\widetilde{X}_{1}}$ and $P_{U_{2}|\widetilde{X}_2}$ of the rate tuples such that $\widebar{R}_{\text{s},j} \geq 0$, (\ref{eq:corrkeyrateTwoenroll}) - (\ref{eq:corrleakagerateTwoenrollupperbound}), (\ref{eq:cscorrstoragerateTwoenroll}) - (\ref{eq:cssumonsamesjandwjandjprimeTwoenroll}), and $U_j-\widetilde{X}_{j}-X-Y_j$ form a Markov chain for $j=1,2$. 
\end{theorem}

The inner and outer bounds differ because the outer bounds define rate regions for the Markov chains $U_1-\widetilde{X}_{1}-X-Y_1$ and $U_2-\widetilde{X}_{2}-X-Y_2$, which are larger than the rate regions defined by the inner bounds that satisfy (\ref{eq:fixedprobdistrTwoenroll}). For instance, in the achievability proof of Theorem~\ref{theo:Twoenrollmentgscs}, we apply the properties of the Markov chain $U_2-\widetilde{X}_2-U_1$ in (\ref{eq:conversJ2thatshowcombinationboundisalreadysatisfied})(b), which does not form a Markov chain for the choice of $U_{1}$ and $U_2$ in the outer bounds. Therefore, inner and outer bounds do not match in general. 

\begin{corollary}
	Choosing $U_1=\widetilde{X}_1$ and $U_2=\widetilde{X}_2$, it is straightforward to show that inner and outer bounds in Theorem~\ref{theo:Twoenrollmentgscs} match if we do not impose any storage or privacy constraints, i.e., impose only (\ref{eq:reliabilityconstJ2}), (\ref{eq:uniformityconstJ2}), and (\ref{eq:secrecyconstJ2}). This result improves on the secret-key capacity region given in \cite[Theorem 1]{bizimBenelux} for a weak secrecy constraint. 
\end{corollary}

\begin{example}\label{ex:example1}
	Consider the RO PUF model from \cite[Section 4.1]{bizimMDPI} where a transform-coding method is applied to conservatively model the measurement channels $P_{Y|X}=P_{\widetilde{X}|X}$ as independent BSCs with the same crossover probability of $p_\text{A}$ and where the hidden source output is $\text{Bern}(\frac{1}{2})$. We therefore can apply the achievability results from Theorem~\ref{theo:Twoenrollmentgscs} to this RO PUF model. Using \cite[Theorem~3]{bizimMMMMTIFS} to evaluate the boundary tuples of $\mathcal{\widebar{R}}_{\text{gs},J=2}$, it suffices to consider probability distributions $P_{U_j|\widetilde{X}_j}$ for $j=1,2$ such that $P_{\widetilde{X}_j|U_j}$ are BSCs with crossover probabilities 
	\begin{align}
		\tilde{x}_j = \frac{H_b^{-1}(H(X|U_j))-p_\text{A}}{1-2p_\text{A}}. \label{eq:optimalcrossoverUtoXtilde}
	\end{align}
	Consider the projection of the boundary tuples of $\mathcal{\widebar{R}}_{\text{gs},J=2}$ onto key-leakage plane, i.e., (\ref{eq:corrkeyrateTwoenroll}) and (\ref{eq:corrleakagerateTwoenroll}). We plot in Fig.~\ref{fig:example1plot} single-enrollment results where the privacy-leakage rate is measured with respect to single helper data and two-enrollment results for the sum rate of the two keys, both for $p_\text{A}=0.06$ \cite{bizimMDPI}. To achieve a total secret-key rate of $I(\widetilde{X}_1;Y_1)=I(\widetilde{X}_2;Y_2)$, the privacy-leakage rate for the two-enrollment model is approximately $13.5\%$ less than the privacy-leakage rate for the single-enrollment model for RO PUFs. The reason for this gain is the information bottleneck problem that arises from (\ref{eq:corrkeyrateTwoenroll}) and (\ref{eq:corrleakagerateTwoenroll}) to find the boundary tuples.
\end{example}

\begin{figure}[t]
	\centering
%
%
%
\begin{tikzpicture}

\begin{axis}[%
width=6.056cm,
height=7cm,
at={(0cm,0cm)},
scale only axis,
xmin=0,
xmax=0.2,
xlabel style={font=\color{white!15!black}},
xlabel={Privacy-leakage Rate $\widebar{R}_{\ell}$ (bits/symbol)},
ymin=0,
ymax=0.6,
ylabel style={font=\color{white!15!black}},
ylabel={Secret-key Rate $\widebar{R}_{\text{s}}$ (bits/symbol)},
axis background/.style={fill=white},
xmajorgrids,
ymajorgrids,
ticklabel style={
	/pgf/number format/.cd,
	fixed,
	/tikz/.cd,
},
xticklabel style={
	/pgf/number format/precision=5,
},
legend style={at={(0.8,0.986)}, legend cell align=left, align=left, draw=white!15!black}
]
\addplot [color=blue, solid, line width=1pt, mark size=2.5pt, mark=o, mark options={solid, blue}]
  table[row sep=crcr]{%
  	0.20	0.491695739424604\\
  	0.19085934142092	0.491695739424604\\
0.18085934142092	0.491695739424604\\
0.169493112488077	0.469077442402172\\
0.159058589995795	0.447275745162471\\
0.149413948471083	0.426248112313884\\
0.140450428466459	0.405956975068997\\
0.132082526209982	0.386368929978218\\
0.124241637450467	0.367454101974137\\
0.116871772992157	0.3491856316753\\
0.109926574971911	0.33153925807412\\
0.103367181329448	0.314492975431987\\
0.0971606612737074	0.298026748609704\\
0.0912788457459605	0.282122274922354\\
0.0856974376419761	0.266762783408955\\
0.080395324300233	0.251932864468852\\
0.0753540389131127	0.237618324353786\\
0.0705573333755378	0.223806060164287\\
0.0659908357398914	0.210483951883675\\
0.0616417727526021	0.197640768664727\\
0.0574987430515138	0.185266087114422\\
0.0535515302279075	0.173350219738392\\
0.0497909475705685	0.161884152036051\\
0.0462087082198366	0.150859486999938\\
0.0427973158739662	0.140268395983691\\
0.0395499722494875	0.130103575073528\\
0.036460498299222	0.12035820623684\\
0.033523266804847	0.111025922635022\\
0.0307331444341403	0.102100777581158\\
0.028085441721446	0.0935772167004902\\
0.0255758697190018	0.0854500529160205\\
0.0232005022953796	0.0777144439353261\\
0.0209557432393565	0.0703658719599315\\
0.0188382974735228	0.0634001253766627\\
0.0168451457997417	0.0568132822227668\\
0.0149735226942226	0.0506016952440697\\
0.0132208967480962	0.0447619783889759\\
0.0115849534135319	0.0392909946012908\\
0.0100635797684336	0.0341858447922458\\
0.00865485105673447	0.0294438578871677\\
0.00735701879804029	0.0250625818553176\\
0.00616850029118265	0.0210397756428602\\
0.00508786936224803	0.0173734019389331\\
0.00411384822970673	0.0140616207136371\\
0.00324530037810078	0.0111027834745769\\
0.00248122434792286	0.00849542819556404\\
0.0018207483633127	0.00623827487734185\\
0.00126312573139153	0.00433022170582831\\
0.00080773095777964	0.00277034177850588\\
0.000454056532364855	0.00155788037428739\\
0.000201710347941586	0.000692252746531841\\
5.04137221097878e-05	0.000173042422952907\\
0	0\\
};
\addlegendentry{Single-enrollment}

\addplot [color=red,  dashed, line width=1pt, mark size=2.5pt, mark=triangle, mark options={solid, red}]
  table[row sep=crcr]{%
0.36171868284184	0.983391478849207\\
0.338986224976154	0.938154884804344\\
0.31811717999159	0.894551490324942\\
0.298827896942166	0.852496224627768\\
0.280900856932918	0.811913950137994\\
0.264165052419964	0.772737859956436\\
0.248483274900933	0.734908203948274\\
0.233743545984315	0.698371263350599\\
0.219853149943821	0.663078516148241\\
0.206734362658897	0.628985950863975\\
0.194321322547415	0.596053497219409\\
0.182557691491921	0.564244549844708\\
0.171394875283952	0.533525566817911\\
0.160790648600466	0.503865728937704\\
0.150708077826225	0.475236648707573\\
0.141114666751076	0.447612120328575\\
0.131981671479783	0.42096790376735\\
0.123283545505204	0.395281537329454\\
0.114997486103028	0.370532174228845\\
0.107103060455815	0.346700439476785\\
0.0995818951411369	0.323768304072102\\
0.0924174164396732	0.301718973999876\\
0.0855946317479324	0.280536791967382\\
0.0790999444989751	0.260207150147056\\
0.0729209965984441	0.240716412473679\\
0.0670465336096939	0.222051845270045\\
0.0614662888682806	0.204201555162316\\
0.056170883442892	0.18715443340098\\
0.0511517394380037	0.170900105832041\\
0.0464010045907592	0.155428887870652\\
0.0419114864787129	0.140731743919863\\
0.0376765949470457	0.126800250753325\\
0.0336902915994834	0.113626564445534\\
0.0299470453884452	0.101203390488139\\
0.0264417934961925	0.0895239567779518\\
0.0231699068270639	0.0785819892025816\\
0.0201271595368673	0.0683716895844917\\
0.0173097021134689	0.0588877157743355\\
0.0147140375960806	0.0501251637106352\\
0.0123370005823653	0.0420795512857204\\
0.0101757387244961	0.0347468038778662\\
0.00822769645941346	0.0281232414272743\\
0.00649060075620156	0.0222055669491539\\
0.00496244869584572	0.0169908563911281\\
0.00364149672662539	0.0124765497546837\\
0.00252625146278307	0.00866044341165662\\
0.00161546191555928	0.00554068355701176\\
0.00090811306472971	0.00311576074857478\\
0.000403420695883172	0.00138450549306368\\
0.000100827444219576	0.000346084845905814\\
0	0\\
};
\addlegendentry{Two-enrollment Sum Rate}

\end{axis}
\end{tikzpicture}%
	\caption{Privacy-leakage vs. secret-key rate projection of the boundary tuples of the single- and two-enrollment RO PUF models with BSCs$(p_\text{A}=0.06)$.} 
	\label{fig:example1plot}
\end{figure}

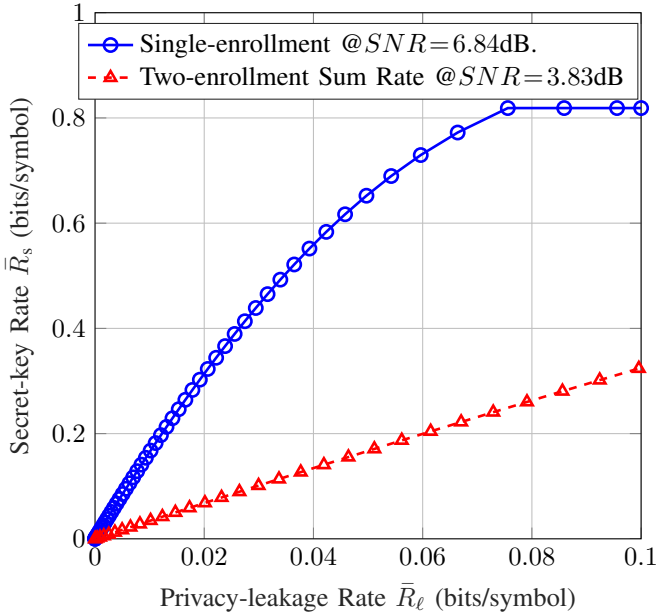
\begin{figure}[t]
	\centering
%
%
%
\begin{tikzpicture}

\begin{axis}[%
width=7.26cm,
height=7cm,
at={(0cm,0cm)},
scale only axis,
xmin=0,
xmax=0.1,
xlabel style={font=\color{white!15!black}},
xlabel={Privacy-leakage Rate $\widebar{R}_{\ell}$ (bits/symbol)},
ymin=0,
ymax=1,
ylabel style={font=\color{white!15!black}},
ylabel={Secret-key Rate $\widebar{R}_{\text{s}}$ (bits/symbol)},
axis background/.style={fill=white},
xmajorgrids,
ymajorgrids,
ticklabel style={
	/pgf/number format/.cd,
	fixed,
	/tikz/.cd,
},
xticklabel style={
	/pgf/number format/precision=4,
},
legend style={at={(0.995,0.9806)},legend cell align=left, align=left, draw=white!15!black}
]
\addplot [color=blue, solid, line width=1pt, mark size=2.5pt, mark=o, mark options={solid, blue}]
table[row sep=crcr]{%
	0.1                             	0.818749563192837\\
	0.0955910764089968	0.818749563192837\\
	0.085910764089968	0.818749563192837\\
0.0755910764089968	0.818749563192837\\
0.0664099290577275	0.772269653505128\\
0.0596469136806566	0.729458970083704\\
0.0542370288271344	0.689585708562623\\
0.0497046709769667	0.652171169687531\\
0.0457932628456846	0.616877602862384\\
0.0423476806660401	0.583453949774301\\
0.0392665900389282	0.551706325935581\\
0.0364799544161121	0.521480562418894\\
0.0339372396419565	0.49265121168118\\
0.0316007230046793	0.465114276130073\\
0.0294414594634309	0.438782205396702\\
0.0274367281975338	0.413580340958088\\
0.0255683501752635	0.389444319751633\\
0.0238215418543182	0.36631813378864\\
0.0221841116155657	0.344152650892966\\
0.0206458824854662	0.322904467304054\\
0.0191982684895973	0.302535004081751\\
0.0178339578787249	0.283009785896325\\
0.0165466723114965	0.264297858475489\\
0.0153309810589646	0.246371312995844\\
0.0141821557507014	0.229204894037919\\
0.0130960554545154	0.212775673608765\\
0.0120690347691729	0.19706277796247\\
0.0110978695981409	0.182047157030574\\
0.0101796966655502	0.167711388552479\\
0.00931196382696542	0.154039510701491\\
0.00849238894317317	0.141016878294016\\
0.00771892560857879	0.128630038658433\\
0.00698973441326278	0.116866624004991\\
0.00630315870781284	0.105715257734998\\
0.00565770405949062	0.0951654725974684\\
0.00505202075592437	0.0852076389743843\\
0.00448488884174758	0.0758329018740098\\
0.00395520527408677	0.0670331254520166\\
0.00346197286157524	0.0588008440750977\\
0.00300429071378594	0.0511292191009389\\
0.00258134597749859	0.0440120006792266\\
0.00219240667589971	0.0374434939865633\\
0.0018368154988474	0.0314185293981917\\
0.00151398441835582	0.0259324361748108\\
0.00122339002475491	0.0209810193062983\\
0.000964569496525125	0.0165605392080377\\
0.000737117131380916	0.0126676940115835\\
0.000540681378354946	0.00929960423101495\\
0.000374962320931038	0.00645379962070725\\
0.000239709570037494	0.00412820807036152\\
0.000134720533263688	0.00232114640977688\\
5.98390332661181e-05	0.00103131301969728\\
1.49542541646452e-05	0.000257782166699028\\
0	0\\
};
\addlegendentry{Single-enrollment @$SNR\!=\!6.84$dB.}

\addplot [color=red,  dashed, line width=1pt, mark size=2.5pt, mark=triangle, mark options={solid, red}]
table[row sep=crcr]{%
0.36171868284184	0.983391478849207\\
0.338986224976154	0.938154884804344\\
0.31811717999159	0.894551490324942\\
0.298827896942166	0.852496224627768\\
0.280900856932918	0.811913950137994\\
0.264165052419964	0.772737859956436\\
0.248483274900933	0.734908203948274\\
0.233743545984315	0.698371263350599\\
0.219853149943821	0.663078516148241\\
0.206734362658897	0.628985950863975\\
0.194321322547415	0.596053497219409\\
0.182557691491921	0.564244549844708\\
0.171394875283952	0.533525566817911\\
0.160790648600466	0.503865728937704\\
0.150708077826225	0.475236648707573\\
0.141114666751076	0.447612120328575\\
0.131981671479783	0.42096790376735\\
0.123283545505204	0.395281537329454\\
0.114997486103028	0.370532174228845\\
0.107103060455815	0.346700439476785\\
0.0995818951411369	0.323768304072102\\
0.0924174164396732	0.301718973999876\\
0.0855946317479324	0.280536791967382\\
0.0790999444989751	0.260207150147056\\
0.0729209965984441	0.240716412473679\\
0.0670465336096939	0.222051845270045\\
0.0614662888682806	0.204201555162316\\
0.056170883442892	0.18715443340098\\
0.0511517394380037	0.170900105832041\\
0.0464010045907592	0.155428887870652\\
0.0419114864787129	0.140731743919863\\
0.0376765949470457	0.126800250753325\\
0.0336902915994834	0.113626564445534\\
0.0299470453884452	0.101203390488139\\
0.0264417934961925	0.0895239567779518\\
0.0231699068270639	0.0785819892025816\\
0.0201271595368673	0.0683716895844917\\
0.0173097021134689	0.0588877157743355\\
0.0147140375960806	0.0501251637106352\\
0.0123370005823653	0.0420795512857204\\
0.0101757387244961	0.0347468038778662\\
0.00822769645941346	0.0281232414272743\\
0.00649060075620156	0.0222055669491539\\
0.00496244869584572	0.0169908563911281\\
0.00364149672662539	0.0124765497546837\\
0.00252625146278307	0.00866044341165662\\
0.00161546191555928	0.00554068355701176\\
0.00090811306472971	0.00311576074857478\\
0.000403420695883172	0.00138450549306368\\
0.000100827444219576	0.000346084845905814\\
0	0\\
};
\addlegendentry{Two-enrollment Sum Rate @$SNR\!=\!3.83$dB}

\end{axis}

\end{tikzpicture}%
	\caption{Privacy-leakage vs. secret-key rate projection of the boundary tuples of the single- and two-enrollment RO PUF models with different SNRs.} 
	\label{fig:Gaussianexample2}
\end{figure}

\begin{example}
	Consider uniform binary antipodal measurements over an additive white Gaussian noise (AWGN) channel. Define the signal power as $P_{\text{S}}$ and the noise power as $P_{\text{N}}$, so we have a signal-to-noise ratio (SNR) of $\displaystyle SNR = \frac{P_{\text{S}}}{P_{\text{N}}}$. If a matched filter, which maximizes the $SNR$ at the sampling instant for the AWGN channel, is applied at the encoder and decoder, the bit error probability $P_{\text{b}}$ is given by \cite[pp. 96]{bizimSingaporeLNs}
	 \begin{align}
	 	P_{\text{b}} = Q\left(\sqrt{SNR}\right).
	 \end{align}  
	 The channel between input binary symbols and outputs of the matched filter is a BISO channel. Using \cite[Theorem~3]{bizimMMMMTIFS}, we have that $P_{\widetilde{X}_j|U_j}$ for $j=1,2$ that are BSCs with crossover probabilities given in (\ref{eq:optimalcrossoverUtoXtilde}) by replacing $p_\text{A}$ with $P_{\text{b}}$, suffice to obtain the boundary tuples of $\mathcal{\widebar{R}}_{\text{gs},J=2}$. We remark that $p_{\text{A}}=0.06$ used in Example~\ref{ex:example1} corresponds to an SNR of approximately $3.83$dB.
	 
	 In Fig.~\ref{fig:Gaussianexample2}, the privacy-leakage rate vs. secret-key rate boundary tuples are depicted for two cases. First, a two-enrollment model at $SNR=3.83$dB with a sum rate for two secret keys is depicted, where each enrollment has a signal power of $P_s$. For comparison, we plot a single-enrollment model with the signal power of $2P_{\text{s}}$, i.e., we have an SNR of approximately $6.84$dB.  Fig.~\ref{fig:Gaussianexample2} shows for the two cases with the same total signal power of $2P_{\text{s}}$, unlike in Example~\ref{ex:example1}, that the single enrollment boundary tuple can result in a gain of approximately $228.55\%$ at its top left corner point in terms of the secret-key rates achieved for a given privacy-leakage rate. For such an AWGN channel with a fixed total signal power; therefore, the single-enrollment model can result in significant gains in terms of achieved secret-key rates as compared to the two-enrollment model for small $\widebar{R}_{\ell}$ values.  
\end{example}

\section{Proof of Theorem~\ref{theo:GeneralInnergscs}}\label{sec:achinnergeneral}
We provide a proof that follows from the output statistics of random binning (OSRB) method, proposed in \cite{OSRBAmin} and further extended in \cite{YenerOSRB}, by applying the steps in  \cite[Section 1.6]{BlochLectureNotes2018}. 

\subsection{Proof for the GS Model}\label{subsec:Theorem1proofGS}

\begin{IEEEproof}[Proof Sketch] Fix $\displaystyle P_{U_1|\widetilde{X}_1}, \displaystyle P_{U_2|\widetilde{X}_2},\ldots,P_{U_J|\widetilde{X}_J}$. Let $(\mathcal{U}_{[1:J]}^n,\mathcal{\widetilde{X}}_{[1:J]}^n,X^n,\mathcal{Y}_{[1:J]}^n)$ be i.i.d. according to (\ref{eq:fixedprobdistr}). Assign three random bin indices $(S_j,W_j,C_j)$ to each realization $u_j^n$ for all $j\in[1:J]$, where $S_j$ represents the secret key, $W_j$ the helper data, and $C_j$ a public index referring to a random encoder-decoder pair fixed below. Assume $S_j\in[1:2^{nR_{\text{s},j}}]$, $W_j\in[1:2^{nR_{\text{w},j}}]$, and $C_j\in[1:2^{nR_{\text{c},j}}]$ such that $R_{\text{s},j},R_{\text{w},j},R_{\text{c},j}\geq0$ for all $j\in[1:J]$.

Apply the union bound to the reliability constraint in (\ref{eq:reliabilityconst}) to obtain the sum of $J$ error probabilities. This sum vanishes for any finite number $J$ when $n\rightarrow\infty$ by using a Slepian-Wolf (SW) \cite{SW} decoder to estimate $U_j^n$ from $(C_j,W_j,Y_j^n)$ if \cite[Lemma 1]{OSRBAmin}
\begin{align}
	R_{\text{c},j}+R_{\text{w},j}> H(U_j|Y_j),\qquad\forall j\in[1:J]. \label{eq:slepianwolfdecoder}
\end{align}
	
The key uniformity (\ref{eq:uniformityconst}), mutual and strong key independence (\ref{eq:keyindependence}), and strong secrecy (\ref{eq:secrecyconst}) constraints are satisfied if \cite[Theorem 1]{OSRBAmin}
\begin{align}
R_{\text{s},j}\!+\!R_{\text{w},j}\!+\!R_{\text{c},j}< H(U_j|\,\mathcal{U}_{[1:J]\setminus \{j\}}), \;\;\;\forall j\in[1:J]\label{eq:independenceofindices}
\end{align}
since (\ref{eq:independenceofindices}) ensures that the three random indices $(S_j,W_j,C_j)$ are almost mutually independent and uniformly distributed, and they are almost independent of $\mathcal{U}_{[1:J]\setminus\{j\}}$. Therefore, $(S_j,W_j,C_j)$ are almost independent of $\left(\mathcal{S}_{[1:J]\setminus\{j\}}, \mathcal{W}_{[1:J]\setminus\{j\}},  \mathcal{C}_{[1:J]\setminus\{j\}}\right)$ because $U_k^n$ determines $(S_k,W_k,C_k)$ for all $k\in[1:J]$.
	
Similarly, the public randomness $C_j$ is almost independent of $\widetilde{X}^n_j$, so it is almost independent of $(\mathcal{\widetilde{X}}_{[1:J]}^n,X^n,\mathcal{Y}_{[1:J]}^n)$, if we have \cite[Theorem 1]{OSRBAmin}
\begin{align}
	R_{\text{c},j}<H(U_j|\widetilde{X}_j),\qquad \qquad \forall j\in[1:J].\label{eq:independenceofcode}
\end{align} 
Thus, the public indices $\mathcal{C}_{[1:J]}$ can be fixed and shared with all parties by generating them uniformly at random. The $j$-th encoder can generate $U_j^n$ according to $P_{U_j^n|\widetilde{X}_j^nC_{j} }$ obtained from the binning scheme above to compute the bins $(S_j,W_{j})$ from $U_j^n$ for all $j\in[1:J]$. This procedure induces a joint probability distribution that is almost equal to $P_{\mathcal{U}_{[1:J]}\mathcal{\widetilde{X}}_{[1:J]}X\mathcal{Y}_{[1:J]}}$ fixed in (\ref{eq:fixedprobdistr}) \cite[Section 1.6]{BlochLectureNotes2018}. 

Applying the Fourier Motzkin elimination \cite{FMEbook} using the software available in \cite{FMEZiv} to (\ref{eq:slepianwolfdecoder})-(\ref{eq:independenceofcode}) for each $j\in[1:J]$ separately, we obtain the inequalities
\begin{align}
&R_{\text{w},j}>I(U_j;\widetilde{X}_j)-I(U_j;Y_j)\label{eq:FMEgeneralinnerrw}\\
&R_{\text{s},j}<I(U_j;Y_j)-I(U_j;\mathcal{U}_{[1:J]\setminus\{j\}})\label{eq:FMEgeneralinnerrs}\\
&R_{\text{w},j}+R_{\text{s},j}< H(U_j|\mathcal{U}_{[1:J]\setminus\{j\}})\label{eq:FMEgeneralinnerrwplusrs}	
\end{align}
for all $j\in[1:J]$.

To satisfy the constraints (\ref{eq:FMEgeneralinnerrw})-(\ref{eq:FMEgeneralinnerrwplusrs}), we can fix the rates to
\begin{alignat}{2}
	&R_{\text{s},j} = I(U_j;Y_j)\!-\!I(U_j;\mathcal{U}_{[1:J]\setminus\{j\}})\!-\!2\epsilon,&& \quad\forall j\in[1:J]\label{eq:chooseR_s}\\
	&R_{\text{w},j} = I(U_j;\widetilde{X}_j)-I(U_j;Y_j)+2\epsilon,&&\quad\forall j\in[1:J]\label{eq:chooseR_w}\\
	&R_{\text{c},j} = H(U_j|\widetilde{X}_j)-\epsilon,&&\quad\forall j\in[1:J]\label{eq:chooseR_c}
\end{alignat}
for some $\epsilon>0$ such that $\epsilon\rightarrow0$ when $n\rightarrow\infty$. 

Consider the privacy leakage. Since $\mathcal{C}_{[1:J]}$ are public, we can bound the privacy leakage as follows.
\begin{align}
	& I(X^n;\mathcal{W}_{[1:J]},\mathcal{C}_{[1:J]})\nonumber\\
	&\quad\leq H(\mathcal{W}_{[1:J]})-H(\mathcal{W}_{[1:J]},\mathcal{C}_{[1:J]}|X^n)+H(\mathcal{C}_{[1:J]})\nonumber\\
	&\quad\overset{(a)}{=}H(\mathcal{W}_{[1:J]})-\sum_{j=1}^JH(W_j,C_j|X^n)+H(\mathcal{C}_{[1:J]})\nonumber\\
	&\quad\leq\sum_{j=1}^{J}\Big(H(W_j)\!+\!H(C_j)\!-\!H(W_j,C_j|X^n)\Big)\label{eq:privacyleakagefirstpartach}
\end{align} 
where $(a)$ follows because $(W_j,C_j)-X^n-(\mathcal{W}_{[1:j-1]},\mathcal{C}_{[1:j-1]})$ form a Markov chain for all $j\in[2:J]$.

Consider two cases for the privacy leakage analysis.

\textbf{Case 1:} Suppose for any $j\in[1:J]$ that we have
\begin{align}
	R_{\text{c},j}+R_{\text{w},j}< H(U_j|X)
\end{align}
i.e., $H(U_j|X)>H(U_j|Y_j)$, so $(W_j,C_j,X^n)$ are almost mutually independent \cite[Theorem 1]{OSRBAmin}. Therefore, we have 
\begin{align}
	&H(W_j)\!+\!H(C_j)\!-\!H(W_j,C_j|X^n)\nonumber\\
	&\quad\leq H(W_j)\!+\!H(C_j)\!-\!(H(W_j)+H(C_j)-\epsilon_n')=\epsilon_n'\label{eq:privleakCase1}
\end{align}
for some $\epsilon_n'>0$ such that $\epsilon_n'\rightarrow 0$ when $n\rightarrow\infty$. Combining (\ref{eq:privacyleakagefirstpartach}) and (\ref{eq:privleakCase1}) proves strong privacy.

\textbf{Case 2:} Suppose for any $j\in[1:J]$ that we have
\begin{align}
R_{\text{c},j}+R_{\text{w},j}\geq H(U_j|X) \label{eq:degradedXtildeXY case2}
\end{align}
i.e., $H(U_j|X)\leq H(U_j|Y_j)$, so $(W_j,C_j,X^n)$ can reliably estimate $U_j^n$ \cite[Lemma 1]{OSRBAmin}. Therefore, we have 
\begin{align}
&H(W_j)\!+\!H(C_j)\!-\!H(W_j,C_j|X^n)\nonumber\\
&\quad\overset{(a)}{\leq}H(W_j)\!+\!H(C_j)\!-\!nH(U_j|X)+n\epsilon_n''\nonumber\\
&\quad\overset{(b)}{\leq}n(I(U_j;X)-I(U_j;Y_j)+\epsilon+\epsilon_n'')\label{eq:privleakCase2}
\end{align}
where $(a)$ follows because $U_j^n$ determines $(W_j,C_j)$, $(W_j,C_j,X^n)$ can realiably estimate $U^n$ for some $\epsilon_n''>0$ such that $\epsilon_n''\rightarrow 0$ when $n\rightarrow\infty$, and $(U_j^n,X^n)$ are i.i.d., and $(b)$ follows by (\ref{eq:chooseR_w}) and (\ref{eq:chooseR_c}).

Combining (\ref{eq:privacyleakagefirstpartach}) and (\ref{eq:privleakCase2}), we obtain
\begin{align}
	&I(X^n;\mathcal{W}_{[1:J]},\mathcal{C}_{[1:J]})\nonumber\\
	&\quad\!\leq \!\sum_{\substack{j=1\\\; j:\\H(U_j|X)\leq H(U_j|Y_j)}}^J\! n(I(U_j;X)\!-\!I(U_j;Y_j)\!+\!\epsilon\!+\!\epsilon_n'').
\end{align}

Using the selection lemma \cite[Lemma 2.2]{Blochbook}, these prove the achievability of the  rate region $\mathcal{R}_{\text{gs}}$.
\end{IEEEproof}

\subsection{Proof for the CS Model}\label{subsec:Theorem1proofCS}
We use the achievability proof for the GS model. Suppose the key $S'_j$, generated as in the GS model together with the helper data $W'_j$ and public index $C'_j$, have the same cardinality as the corresponding embedded secret key $S_j$, i.e., $|\mathcal{S}'_j|=|\mathcal{S}_j|$ for all $j\in [1:J]$. The chosen key $S_j$ is uniformly distributed and independent of $(X^n,\mathcal{\widetilde{X}}_{[1:J]}^n,\mathcal{Y}_{[1:J]}^n, \mathcal{S}_{[1:J]\setminus\{j\}})$ for all $j\in[1:J]$. Consider the $j$-th encoder $f_{\text{cs},j}(\cdot,\cdot)$ with inputs $(\widetilde{X}^n_j,S_j)$ and output $W_j=(S'_j+S_j,W'_j)$, and the $j$-th decoder $g_j(\cdot,\cdot)$ with inputs $(Y_j^n,W_j)$ and output $\hat{S}_j=S'_j+S_j-\hat{S}'_j$. All addition and subtraction operations are modulo-$|\mathcal{S}_j|$ for all $j\in[1:J]$. The $j$-th decoder of the GS model is used to obtain $\hat{S}'_j$ for all $j\in[1:J]$. 

We have the error probability
\begin{align}
&\Pr\left[\underset{j\in[1:J]}{\bigcup}\{S_j\ne\hat{S}_j\}\right]=\Pr\left[\underset{j\in[1:J]}{\bigcup}\{S'_j\ne\hat{S}'_j\}\right]\label{eq:errorprobabilityachtheo2}
\end{align}
which is small due to the proof of achievability for the GS model.

Using (\ref{eq:chooseR_s}) and (\ref{eq:chooseR_w}), and from the one-time padding operation applied above, we can achieve a storage rate of
\begin{align}
	&R_{\text{w},j} \geq I(U_j;\widetilde{X}_j) - I(U_j;U_{[1:J]\setminus\{j\}}),\quad\; \forall j\in[1:J]\label{eq:csstorageach}
\end{align}
for the CS model.

We have the secrecy leakage of
\begin{align}
	&I(\mathcal{S}_{[1:J]};\mathcal{W}_{[1:J]},\mathcal{C}_{[1:J]}')\overset{(a)}{=}I(\mathcal{S}_{[1:J]};\mathcal{W}_{[1:J]}|\mathcal{C}_{[1:J]}')\nonumber\\
	&\!=\! I(\mathcal{S}_{[1:J]};\mathcal{W}_{[1:J]}'|\mathcal{C}_{[1:J]}'\!)\!+\!I(\mathcal{S}_{[1:J]};{(\mathcal{S}'\!+\!\mathcal{S})}_{[1:J]}\!|\mathcal{W}_{[1:J]}',\mathcal{C}_{[1:J]}'\!)\nonumber\\
	&\!\overset{(b)}{=} H({(\mathcal{S}'\!+\!\mathcal{S})}_{[1:J]}|\mathcal{W}_{[1:J]}',\mathcal{C}_{[1:J]}') - H(\mathcal{S}_{[1:J]}'|\mathcal{W}_{[1:J]}',\mathcal{C}_{[1:J]}') \nonumber\\
	&\!\overset{(c)}{\leq} n\Big(\sum_{j=1}^JR_{\text{s},j}\Big)-H(\mathcal{S}_{[1:J]}'|\mathcal{C}_{[1:J]}')+I(\mathcal{S}_{[1:J]}';\mathcal{W}_{[1:J]}'|\mathcal{C}_{[1:J]}')\nonumber\\
	&\!\overset{(d)}{\leq} n\Big(\sum_{j=1}^JR_{\text{s},j}\Big)-\Big(n\Big(\sum_{j=1}^JR_{\text{s},j}\Big)-\epsilon_n'''\Big)\nonumber\\
	&\qquad+I(\mathcal{S}_{[1:J]}';\mathcal{W}_{[1:J]}'|\mathcal{C}_{[1:J]}')\nonumber\\
	&\!\overset{(e)}{\leq}\epsilon_n'''+\epsilon_n^{(4)}
\end{align}
where $(a)$ follows since $\mathcal{S}_{[1:J]}$ are chosen independently of the public indices $\mathcal{C}_{[1:J]}$, $(b)$ follows because $\mathcal{S}_{[1:J]}$ are chosen independently of $(\mathcal{W}_{[1:J]}',\mathcal{C}_{[1:J]}',\mathcal{S}_{[1:J]}')$, $(c)$ follows because $|\mathcal{S}'_j|=|\mathcal{S}_j|$ for all $j\in[1:J]$, $(d)$ follows because $\mathcal{S}_{[1:J]}'$ and $\mathcal{C}_{[1:J]}'$ are almost mutually independent and each $S_j'$ is almost uniformly distributed due to (\ref{eq:independenceofindices}) for some $\epsilon_n'''>0$ such that $\epsilon_n'''\rightarrow0$ when $n\rightarrow\infty$, and $(e)$ follows because the GS model satisfies the strong secrecy constraint (\ref{eq:secrecyconst}) due to (\ref{eq:independenceofindices}) for some $\epsilon_n^{(4)}>0$ such that $\epsilon_n^{(4)}\rightarrow0$ when $n\rightarrow\infty$.

Consider the privacy leakage:
\begin{align}
&I(X^n;\mathcal{W}_{[1:J]},\mathcal{C}_{[1:J]}')\nonumber\\
	&\leq I(X^n;\mathcal{W}_{[1:J]}',\mathcal{C}_{[1:J]}') +H((\mathcal{S}+\mathcal{S}')_{[1:J]}|\mathcal{W}_{[1:J]}',\mathcal{C}_{[1:J]}')\nonumber\\
	&\qquad -H((\mathcal{S}+\mathcal{S}')_{[1:J]}|X^n,\mathcal{W}_{[1:J]}',\mathcal{C}_{[1:J]}',\mathcal{S}_{[1:J]}')\nonumber\\
	&\overset{(a)}{\leq} I(X^n;\mathcal{W}_{[1:J]}',\mathcal{C}_{[1:J]}')\!+\!\Big(\sum_{j=1}^J\log (|\mathcal{S}_j|)\Big)\!-\!H(\mathcal{S}_{[1:J]})\nonumber\\
	&\overset{(b)}{=}  I(X^n;\mathcal{W}_{[1:J]}',\mathcal{C}_{[1:J]}') \label{eq:ach2privleaktemp}
\end{align}
where $(a)$ follows because $\mathcal{S}_{[1:J]}$ are chosen independently of $(X^n,\mathcal{W}_{[1:J]}',\mathcal{S}_{[1:J]}',\mathcal{C}_{[1:J]}')$ and $|\mathcal{S}'_j|=|\mathcal{S}_j|$ for all $j\in[1:J]$ and $(b)$ follows from the uniformity and mutual independence of $\mathcal{S}_{[1:J]}$. 

Using the selection lemma, these prove the achievability of the  rate region $\mathcal{R}_{\text{cs}}$.

\section{Proof of Thorem~\ref{theo:Twoenrollmentgscs}}\label{sec:Twoenrollmentproofs}
We use the OSRB method steps in \cite[Section 1.6]{BlochLectureNotes2018}. 

\subsection{Achievability Proof for the GS Model}
Fix 
\begin{align}
	\displaystyle P_{U_1|\widetilde{X}_1}=P_{U_2|\widetilde{X}_2}=P_{U|\widetilde{X}}.\label{eq:twoenrolleaulencoder}
\end{align}
Let $(U_{1}^n,U_2^n,\widetilde{X}_{1}^n,\widetilde{X}_{2}^n,X^n,Y_{1}^n,Y_{2}^n)$ be i.i.d. according to (\ref{eq:fixedprobdistrTwoenroll}). Assign three random bin indices $(S_j,W_j,C_j)$ to each realization $u_j^n$ for all $j=1,2$. Assume $S_j\in[1:2^{n\widebar{R}_{\text{s},j}}]$, $W_j\in[1:2^{n\widebar{R}_{\text{w},j}}]$, and $C_j\in[1:2^{n\widebar{R}_{\text{c},j}}]$ such that $\widebar{R}_{\text{s},j},\widebar{R}_{\text{w},j},\widebar{R}_{\text{c},j}\geq0$ for $j=1,2$.

Apply the union bound to the reliability constraint in (\ref{eq:reliabilityconstJ2}), which vanishes when $n\rightarrow\infty$ by using an SW decoder to estimate $U_j^n$ from $(C_j,W_j,Y_j^n)$ if \cite[Lemma 1]{OSRBAmin}
\begin{align}
\widebar{R}_{\text{c},j}+\widebar{R}_{\text{w},j}> H(U_j|Y_j),\qquad\;\;\;\; j=1,2. \label{eq:slepianwolfdecoderTwoenroll}
\end{align}

The key uniformity (\ref{eq:uniformityconstJ2}) constraint is satisfied if \cite[Theorem 1]{OSRBAmin}
\begin{align}
\widebar{R}_{\text{s},j}\!+\!\widebar{R}_{\text{w},j}\!+\!\widebar{R}_{\text{c},j}< H(U_j), \qquad j=1,2\label{eq:keyuniformitytwoenroll}
\end{align}
since (\ref{eq:keyuniformitytwoenroll}) ensures that the three random indices $(S_j,W_j,C_j)$ are almost mutually independent and uniformly distributed.

Suppose a virtual joint encoder assigns six indices $(S_1,W_1,C_1,S_2,W_2,C_2)$ to each realization pair 
$(u_1^n,u_2^n)$. This virtual encoder is an operational dual of the virtual decoder used in the proof of \cite[Theorem 1]{bizimBenelux}. Using the virtual joint encoder, the strong secrecy constraint in (\ref{eq:secrecyconstJ2})  and the strong helper data independence constraint in (\ref{eq:storageindependence}) are satisfied if \cite[Theorem 1]{OSRBAmin}
\begin{align}
\widebar{R}_{\text{s},1}\!+\!\widebar{R}_{\text{w},1}\!+\!\widebar{R}_{\text{c},1}\!+\!\widebar{R}_{\text{w},2}\!+\!\widebar{R}_{\text{c},2}< H(U_1,U_2)\label{eq:keyleakageperkeytwoenroll_part1}
\end{align}
and
\begin{align}
\widebar{R}_{\text{s},2}\!+\!\widebar{R}_{\text{w},2}\!+\!\widebar{R}_{\text{c},2}\!+\!\widebar{R}_{\text{w},1}\!+\!\widebar{R}_{\text{c},1}< H(U_1,U_2)\label{eq:keyleakageperkeytwoenroll_part2}
\end{align}
because (\ref{eq:keyleakageperkeytwoenroll_part1}) ensures that $(S_1,W_1,C_1,W_2,C_2)$ are almost mutually independent; whereas, (\ref{eq:keyleakageperkeytwoenroll_part2}) ensures that $(S_2,W_2,C_2,W_1,C_1)$ are almost mutually independent.

\begin{remark}\label{rem:whylessstringent}
	\normalfont The set of equations considered in (\ref{eq:keyuniformitytwoenroll})-(\ref{eq:keyleakageperkeytwoenroll_part2}) cannot be imposed for the joint secrecy-leakage constraint in (\ref{eq:secrecyconst}) for general probability distributions $P_{\widetilde{X}_1\widetilde{X}_2XY_1Y_2}$, since to impose (\ref{eq:secrecyconst}) one would replace  (\ref{eq:keyleakageperkeytwoenroll_part1}) and (\ref{eq:keyleakageperkeytwoenroll_part2}) with
	\begin{align}
		\widebar{R}_{\text{s},1}\!+\!\widebar{R}_{\text{w},1}\!+\!\widebar{R}_{\text{c},1}\!+	\widebar{R}_{\text{s},2}+\!\widebar{R}_{\text{w},2}\!+\!\widebar{R}_{\text{c},2}< H(U_1,U_2)\label{eq:SStringentkeyleakageperkeytwoenroll_part1}
	\end{align}
	which would also imply the mutual independence of secret keys in (\ref{eq:keyindependence}). However, the inequalities in (\ref{eq:keyuniformitytwoenroll}) and (\ref{eq:SStringentkeyleakageperkeytwoenroll_part1}) cannot be satisfied simultaneously in general as $H(U_1)+H(U_2)\geq H(U_1,U_2)$. This problem is avoided in the proof of Theorem~\ref{theo:GeneralInnergscs} by imposing the inequality in (\ref{eq:independenceofindices}) rather than (\ref{eq:keyuniformitytwoenroll}).
\end{remark}

The public randomness $C_j$ is almost independent of $\widetilde{X}^n_j$, so it is almost independent of $(\widetilde{X}_1^n,\widetilde{X}_2^n,X^n,Y_1^n,Y_2^n)$, if we have \cite[Theorem 1]{OSRBAmin}
\begin{align}
\widebar{R}_{\text{c},j}<H(U_j|\widetilde{X}_j),\qquad \qquad  j=1,2.\label{eq:independenceofcodetwoenroll}
\end{align} 
Thus, the public indices $(C_1,C_2)$ can be fixed and shared publicly by generating them uniformly at random. $U_j^n$ can be generated according to $P_{U_j^n|\widetilde{X}_j^nC_j}$ for $j=1,2$ obtained from the binning scheme above to compute the bins $(S_j,W_{j})$ from $U_j^n$ for $j=1,2$. This procedure induces a joint probability distribution that is almost equal to $P_{U_{1}U_2\widetilde{X}_{1}\widetilde{X}_2XY_{1}Y_2}$ that is fixed in (\ref{eq:fixedprobdistrTwoenroll}) \cite[Section 1.6]{BlochLectureNotes2018}. 

Applying the Fourier Motzkin elimination to (\ref{eq:slepianwolfdecoderTwoenroll})-(\ref{eq:keyleakageperkeytwoenroll_part2}) and (\ref{eq:independenceofcodetwoenroll}), we obtain the inequalities
\begin{align}
	&\widebar{R}_{\text{w},1}> H(U_1|Y_1)-H(U_1|\widetilde{X}_1)\label{eq:J2constraints1}\\
	&\widebar{R}_{\text{w},2}> H(U_2|Y_2)-H(U_2|\widetilde{X}_2)\\
	&\widebar{R}_{\text{s},1}<I(U_1;Y_1)\label{eq:J2R1ineq}\\
	&\widebar{R}_{\text{s},2}<I(U_2;Y_2)\label{eq:J2R2ineq}\\
	&\widebar{R}_{\text{s},1}<-H(U_1|Y_1)-H(U_2|Y_2)+H(U_1,U_2)\label{eq:J2unnecessaryRs1}\\
	&\widebar{R}_{\text{s},2}<-H(U_1|Y_1)-H(U_2|Y_2)+H(U_1,U_2)\label{eq:J2unnecessaryRs2}\\
	& \widebar{R}_{\text{s},1} + \widebar{R}_{\text{w},2}<-H(U_1|Y_1)+H(U_1,U_2)\label{eq:J2S1W2}\\
	& \widebar{R}_{\text{s},1} + \widebar{R}_{\text{w},1}<H(U_1)\label{eq:HU1}\\
	& \widebar{R}_{\text{s},1} + \widebar{R}_{\text{w},1}<-H(U_2|Y_2)+H(U_1,U_2)\label{eq:HU1icingereksiz}\\
	& \widebar{R}_{\text{s},1} + \widebar{R}_{\text{w},1}+\widebar{R}_{\text{w},2}<H(U_1,U_2)\label{eq:J2Rs1Rw1Rw2}\\
	& \widebar{R}_{\text{s},2} + \widebar{R}_{\text{w},2}<-H(U_1|Y_1)+H(U_1,U_2)\label{eq:HU2icingereksiz}\\
	& \widebar{R}_{\text{s},2} + \widebar{R}_{\text{w},2}<H(U_2)\label{eq:HU2}\\
	& \widebar{R}_{\text{s},2} + \widebar{R}_{\text{w},1}<-H(U_2|Y_2)+H(U_1,U_2)\label{eq:J2S2W1}\\
	& \widebar{R}_{\text{s},2} + \widebar{R}_{\text{w},2}+\widebar{R}_{\text{w},1}<H(U_1,U_2).\label{eq:J2constraintslast}
\end{align}

Observe that we have
\begin{align}
	&H(U_1|\widetilde{X}_2) = H(U_1|Y_1) = H(U_2|\widetilde{X}_1)=H(U_2|Y_2)\label{eq:uconditionalequalitiesJ2}\\
	& H(U_1|\widetilde{X}_1) = H(U_2|\widetilde{X}_2)\label{eq:u1xtilde1u2xtilde2ineqJ2}\\
	& H(U_1) = H(U_2)\label{eq:u1equaltou2J2}
\end{align}
due to (\ref{eq:SRAMchanneldef}) and (\ref{eq:twoenrolleaulencoder}). We therefore obtain
\begin{align}
	&H(U_1,U_2)-H(U_1|Y_1)\overset{(a)}{=}H(U_2)\!+\!H(U_1|U_2)\!-\!H(U_1|\widetilde{X}_2)\nonumber\\
	&\quad\overset{(b)}{\geq} H(U_2)\label{eq:conversJ2thatshowcombinationboundisalreadysatisfied}
\end{align}
where $(a)$ follows by (\ref{eq:uconditionalequalitiesJ2}) and $(b)$ follows from the Markov chain $U_2-\widetilde{X}_2-U_1$. A similar result can be shown by swaping the indices. Therefore, the constraints in (\ref{eq:HU1icingereksiz}) and (\ref{eq:HU2icingereksiz}) are inactive due to the constraints, respectively, in (\ref{eq:HU1}) and (\ref{eq:HU2}). Similarly, the constraints in (\ref{eq:J2unnecessaryRs1}) and (\ref{eq:J2unnecessaryRs2}) are inactive due to the constraints, respectively, in (\ref{eq:J2R1ineq}) and (\ref{eq:J2R2ineq}).

Replace the inequalities in (\ref{eq:J2S1W2}) and (\ref{eq:J2S2W1}), respectively, with
\begin{align}
&2\widebar{R}_{\text{s},1}+\widebar{R}_{\text{w},1}+\widebar{R}_{\text{w},2}< I(U_1;Y_1)+H(U_1,U_2)\label{eq:intermediateineq3}\\
&2\widebar{R}_{\text{s},2}+\widebar{R}_{\text{w},2}+\widebar{R}_{\text{w},1}< I(U_2;Y_2)+H(U_1,U_2)\label{eq:intermediateineq3forj2}.
\end{align}
Then, (\ref{eq:intermediateineq3}) is inactive because (\ref{eq:J2R1ineq}) and (\ref{eq:J2Rs1Rw1Rw2}) imply (\ref{eq:intermediateineq3}), and (\ref{eq:intermediateineq3forj2}) is inactive because (\ref{eq:J2R2ineq}) and (\ref{eq:J2constraintslast}) imply (\ref{eq:intermediateineq3forj2}). We remark that the rate region represented by (\ref{eq:J2constraints1})-(\ref{eq:J2constraintslast}) is the same as the region represented by replacing (\ref{eq:J2S1W2}) and (\ref{eq:J2S2W1}) with (\ref{eq:intermediateineq3}) and (\ref{eq:intermediateineq3forj2}) because the corner points (i.e., the points that asymptotically achieve equalities in the given inequalities for fixed $P_{U_1|\widetilde{X}_1}=P_{U_2|\widetilde{X}_2}$) of the two rate regions are the same. Therefore, the inequalities in (\ref{eq:J2S1W2}) and (\ref{eq:J2S2W1}) are inactive.

To satisfy the constraints (\ref{eq:J2constraints1})-(\ref{eq:J2constraintslast}), we can fix the rates to
\begin{alignat}{2}
&\widebar{R}_{\text{s},j} = I(U_j;Y_j)-\!5\epsilon,&& \qquad j=1,2\label{eq:chooseR_stwoenroll}\\
&\widebar{R}_{\text{w},j} = I(U_j;\widetilde{X}_j)-I(U_j;Y_j)+2\epsilon,&&\qquad j=1,2\label{eq:chooseR_wtwoenroll}\\
&\widebar{R}_{\text{c},j} = H(U_j|\widetilde{X}_j)-\epsilon,&&\qquad j=1,2\label{eq:chooseR_ctwoenroll}
\end{alignat}
for some $\epsilon>0$ such that $\epsilon\rightarrow0$ when $n\rightarrow\infty$ due to (\ref{eq:uconditionalequalitiesJ2})-(\ref{eq:conversJ2thatshowcombinationboundisalreadysatisfied}).

Since $C_1$ and $C_2$ are public, we can bound the privacy leakage as follows.
\begin{align}
& I(X^n;W_1,W_2,C_1,C_2)\nonumber\\
&\overset{(a)}{\leq}H(W_1,W_2) - H(W_1,C_1|X^n)- H(W_2,C_2|X^n)\nonumber\\
&\qquad + H(C_1,C_2)\nonumber\\
&\overset{(b)}{\leq} H(W_1)+H(W_2)-H(U_1^n|X^n)-H(U_2^n|X^n)+2n\epsilon_n''\nonumber\\
&\qquad +H(C_1)+H(C_2)\label{eq:privacyleakachforfixedratesintermediatestep}\\
&\overset{(c)}{\leq} n(I(U_1;X)-I(U_1;Y_1)+I(U_2;X)-I(U_2;Y_2))\nonumber\\
&\qquad +2n\epsilon_n''+2n\epsilon \label{eq:privacyleakachforfixedrates}
\end{align} 
where $(a)$ follows because $(W_1,C_1)-X^n-(W_2,C_2)$ form a Markov chain, $(b)$ follows for some $\epsilon_n''>0$ such that $\epsilon_n''\rightarrow 0$ when $n\rightarrow\infty$ because for the two-enrollment model considered, (\ref{eq:degradedXtildeXY case2}) is satisfied due to the Markov chain $U_j-X-Y_j$ for $j=1,2$, and $(c)$ follows by (\ref{eq:chooseR_wtwoenroll}) and (\ref{eq:chooseR_ctwoenroll}), and because $(U_1^n,U_2^n,X^n)$ are i.i.d.

Using (\ref{eq:privacyleakachforfixedratesintermediatestep}) for general rate tuples that satisfy the constraints (\ref{eq:J2constraints1})-(\ref{eq:J2constraintslast}), i.e., not only (\ref{eq:chooseR_stwoenroll})-(\ref{eq:chooseR_ctwoenroll}), we can bound the privacy leakage alternatively as
\begin{align}
& I(X^n;W_1,W_2,C_1,C_2)\nonumber\\
&\overset{(a)}{\leq} n\widebar{R}_{\text{w},1}+n\widebar{R}_{\text{w},2}+nI(U_1;X)-nI(U_1;\widetilde{X}_1)\nonumber\\
&\qquad+nI(U_2;X)-nI(U_2;\widetilde{X}_2)+2n\epsilon_n'' \label{eq:privacyleakachforGeneralTuples}
\end{align} 
where $(a)$ follows by (\ref{eq:chooseR_ctwoenroll}) and because $(U_1^n,U_2^n,X^n)$ are i.i.d.

Using the selection lemma, these prove the achievability of the  key-leakage-storage region $\mathcal{\widebar{R}}_{\text{gs},J=2}$.
\subsection{Achievability Proof for the CS Model}
The achievability proof for the CS model follows by applying the one-time padding step used in Section~\ref{subsec:Theorem1proofCS}.

\subsection{Outer Bound Proofs for the Multi-enrollment Models}
Suppose for some $\delta_n\!>\!0$ and $n$, there is a pair of encoders and decoders such that (\ref{eq:reliabilityconstJ2})-(\ref{eq:storageindependence}) are satisfied by some key-leakage-storage tuple $(\widebar{R}_{\text{s},1}, \widebar{R}_{\text{s},2}, \widebar{R}_{\ell}, \widebar{R}_{\text{w},1}, \widebar{R}_{\text{w},2})$. Using (\ref{eq:reliabilityconstJ2}) and Fano's inequality, we obtain
\begin{align}
H(S_j|W_j,Y_j^n)\!\overset{(a)}{\leq}\!H(S_j|\hat{S}_j)\!\leq\!n\epsilon_n,\qquad j=1,2 \label{eq:fanoappJ2} 
\end{align}
where $(a)$ permits randomized decoding, $\epsilon_n\!=\!\delta_n \max\{\widebar{R}_{\text{s},1},\widebar{R}_{\text{s},2}\} \!+\!H_b(\delta_n)/n$ such that $\epsilon_n\!\rightarrow\!0$ if $\delta_n\!\rightarrow\!0$. 

Let $U_{j,i}\triangleq (S_j,W_j,X^{i-1})$, which satisfies the Markov chain $U_{j,i}-\widetilde{X}_{j,i}-X_i-Y_{j,i}$ for all $i\in[1:n]$ and $j=1,2$.

\begin{remark}\label{rem:Markovchainsdiff}
For the choice of $U_{j,i}= (S_j,W_j,X^{i-1})$ (and similarly for $U_{j,i}= (S_j,W_j,Y_j^{i-1})$) for $j\!=\!1,2$, $U_{1,i}-\widetilde{X}_{1,i}-U_{2,i}$ do not form a Markov chain for all $i\in[1:n]$ although for the inner bound we use this Markov chain. This is the reason why inner and outer bounds do not match in general.
\end{remark}

\emph{Proof for (\ref{eq:corrkeyrateTwoenroll})}: We obtain for the multi-enrollment GS and CS models for $j=1,2$ that
\begin{align}
&n(\widebar{R}_{\text{s},j}-\delta_n)\overset{(a)}{\leq} H(S_j)-H(S_j|W_j,Y_j^n)+n\epsilon_n\nonumber\\
&\overset{(b)}{\leq} I(S_j;Y_j^n|W_j)+n\epsilon_n+\delta_n\nonumber\\
&\leq \sum_{i=1}^n \Big[I(S_j,W_j,Y_j^{i-1};Y_{j,i})+\epsilon_n+\frac{\delta_n}{n}\Big]\nonumber\\
&\overset{(c)}{\leq}\sum_{i=1}^n \Big[I(S_j,W_j,X^{i-1};Y_{j,i})+\epsilon_n+\frac{\delta_n}{n}\Big]\nonumber\\
&\overset{(d)}{=}\sum_{i=1}^n \Big[I(U_{j,i};Y_{j,i})+\epsilon_n+\frac{\delta_n}{n}\Big]\label{eq:secretkeyconv1}
\end{align}
$(a)$ follows by (\ref{eq:uniformityconstJ2}) and (\ref{eq:fanoappJ2}), $(b)$ follows by (\ref{eq:secrecyconstJ2}), $(c)$ follows by applying the data-processing inequality to the Markov chain
\begin{align}
 Y_j^{i-1}-(W_j,S_j,X^{i-1})-Y_{j,i},\quad j=1,2,\;\;\forall i\in[1\!:\!n]\label{eq:MarkovchainYiYiminus1}
\end{align}
and $(d)$ follows from the definition of $U_{j,i}$.

\emph{Proof for (\ref{eq:corrleakagerateTwoenroll})}: Observe for the multi-enrollment models that
\begin{align}
&n(\widebar{R}_\ell+\delta_n)\overset{(a)}{=} H(W_1,W_2)-H(W_1|X^n)-H(W_2|X^n)\nonumber\\
&\overset{(b)}{=}H(W_1|Y_1^n)-H(W_1|X^n)+H(W_2|Y_2^n)-H(W_2|X^n)\nonumber\\
&\quad +I(W_1;\widetilde{X}_2^n)+I(W_2;Y_2^n)-I(W_1;W_2)\nonumber\\
&\overset{(c)}{\geq}\sum_{j=1}^2\Big[H(W_j|Y_j^n)-H(W_j|X^n)\Big]\nonumber\\
&\geq \sum_{j=1}^2 \Big[H(S_j,W_j,Y_j^n)-H(S_j|W_j,Y_j^n)-H(Y_j^n)\nonumber\\
&\qquad\qquad-H(S_j,W_j|X^n)\Big]\nonumber\\
&\overset{(d)}{\geq} \sum_{j=1}^2\Big[I(S_j,W_j;X^n)-I(S_j,W_j;Y_j^n)-n\epsilon_n\Big]\nonumber\\
&\overset{(e)}{\geq}\!\sum_{j=1}^2\!\sum_{i=1}^{n}\!\Big[I(S_j,W_j,X^{i-1};X_i) \!-\!I(S_j,W_j,X^{i-1};Y_{j,i})	\!-\!\epsilon_n\Big]\nonumber\\
&\overset{(f)}{=}\!\sum_{j=1}^2\sum_{i=1}^{n}\Big[I(U_{j,i};X_i) -I(U_{j,i};Y_{j,i})	-\epsilon_n\Big]\label{eq:convprivacylowerbound}
\end{align}
where $(a)$ follows by (\ref{eq:privacyconstJ2}) and from the Markov chain $W_1-X^n-W_2$, $(b)$ follows because $I(W_1;Y_1^n)=I(W_1;\widetilde{X}_2^n)$ due to (\ref{eq:SRAMchanneldef}), $(c)$ follows from the Markov chain $W_1-\widetilde{X}_2^n-W_2$, $(d)$ follows by (\ref{eq:fanoappJ2}), $(e)$ follows because the channel and source are memoryless and from the Markov chain in (\ref{eq:MarkovchainYiYiminus1}), and $(f)$ follows from the definition of $U_{j,i}$.

\emph{Proof for (\ref{eq:corrleakagerateTwoenrollupperbound})}: Observe for the multi-enrollment models that
\begin{align}
&n(\widebar{R}_\ell+\delta_n)\!\overset{(a)}{\leq}\! H(W_1)\!+\!H(W_2)\!-\!H(W_1|X^n)\!-\!H(W_2|X^n)\nonumber\\
&\overset{(b)}{\leq} \sum_{j=1}^2\Big[n\widebar{R}_{\text{w},j}+H(S_j,W_j|\widetilde{X}_j^n)-H(S_j,W_j|X^n)+n\epsilon_n\Big]\nonumber
\end{align}
\begin{align}
&\overset{(c)}{=}\sum_{j=1}^2\Big[n\widebar{R}_{\text{w},j}+\sum_{i=1}^nI(S_j,W_j,X^{i-1};X_i)\nonumber\\
&\qquad\qquad-\sum_{i=1}^nI(S_j,W_j,\widetilde{X}_{j}^{i-1};\widetilde{X}_{j,i})+n\epsilon_n\Big]\nonumber\\
&\overset{(d)}{\leq}\sum_{j=1}^2\Big[n\widebar{R}_{\text{w},j}+\sum_{i=1}^nI(S_j,W_j,X^{i-1};X_i)\nonumber\\
&\qquad\qquad-\sum_{i=1}^nI(S_j,W_j,X^{i-1};\widetilde{X}_{j,i})+n\epsilon_n\Big]\nonumber\\
&\overset{(e)}{\leq}\sum_{j=1}^2\Big[n\widebar{R}_{\text{w},j}+\sum_{i=1}^n(I(U_{j,i};X_i)-I(U_{j,i};\widetilde{X}_{j,i}))\nonumber\\
&\qquad\qquad+n\epsilon_n\Big]\label{eq:converseprivleakupperbound}
\end{align}
where $(a)$ follows by (\ref{eq:privacyconstJ2}) and from the Markov chain $W_1-X^n-W_2$, $(b)$ follows by (\ref{eq:fanoappJ2}) and from the Markov chain $S_j-(W_j,X^n)-Y^n$ for $j=1,2$, $(c)$ follows because the channel and source are memoryless, $(d)$ follows from the Markov chain
\begin{align}
X^{i-1}\!-\!(W_j, S_j,\widetilde{X}_j^{i-1})\!-\!\widetilde{X}_{j,i},\;\;\; j=1,2,\;\;\forall i\in[1\!:\!n]\label{eq:MarkovchainXtildeiXiminus1} 
\end{align}
and $(e)$ follows from the definition of $U_{j,i}$.

\emph{Proof for (\ref{eq:corrstoragerateTwoenroll})}: Observe for the multi-enrollment  GS model for $j=1,2$ that 
\begin{align}
&n(\widebar{R}_{\text{w},j}+\delta_n)\overset{(a)}{\geq} H(W_j|Y_j^n)+I(W_j;Y_j^n)\nonumber\\
&\overset{(b)}{\geq} H(S_j,W_j,Y_j^n)-H(Y_j^n)-H(S_j|W_j,Y_j^n)\nonumber\\
&\qquad -H(S_j,W_j|\widetilde{X}_j^n)+I(W_j;Y_j^n)\nonumber\\
&\overset{(c)}{\geq} I(S_j,W_j;\widetilde{X}_j^n)-I(S_j,W_j;Y_j^n)-n\epsilon_{n}\nonumber\\
&\overset{(d)}{=}\!\sum_{i=1}^{n}[I(S_j,W_j,\widetilde{X}_j^{i-1};\widetilde{X}_{j,i})\!-\!I(S_j,W_j,Y_j^{i-1};Y_{j,i})\!-\!n\epsilon_{n}]\nonumber\\
&\overset{(e)}{\geq}\sum_{i=1}^{n}[I(S_j,W_j,X^{i-1};\widetilde{X}_{j,i})\!-\!I(S_j,W_j,X^{i-1};Y_{j,i})\!-\!n\epsilon_{n}]\nonumber\\
&\overset{(f)}{=}\sum_{i=1}^{n}[I(U_{j,i};\widetilde{X}_{j,i})\!-\!I(U_{j,i};Y_{j,i})\!-\!n\epsilon_{n}]\label{eq:storageconv1}
\end{align}
where $(a)$ follows by (\ref{eq:storageconstJ2}), $(b)$ follows from the encoding steps, $(c)$ follows by (\ref{eq:fanoappJ2}), $(d)$ follows because the source and channel are memoryless, $(e)$ follows from the data-processing inequality applied to the Markov chains in (\ref{eq:MarkovchainYiYiminus1}) and
(\ref{eq:MarkovchainXtildeiXiminus1}), and $(f)$ follows from the definition of $U_{j,i}$.

\emph{Proof for (\ref{eq:cscorrstoragerateTwoenroll})}: Observe for the multi-enrollment CS model for $j=1,2$ that 
\begin{align}
&n(\widebar{R}_{\text{w},j}\!+\!\delta_n)\overset{(a)}{\geq}\! I(S_j,W_j;\widetilde{X}_j^n)\!-\!H(S_j|W_j)\!+\!H(S_j,W_j|\widetilde{X}_j^n)\nonumber\\
&\overset{(b)}{\geq} I(S_j,W_j;\widetilde{X}_j^n)+I(S_j;W_j)\overset{(c)}{\geq}\!\sum_{i=1}^{n}I(S_j,W_j,\widetilde{X}_j^{i-1};\widetilde{X}_{j,i})\nonumber\\
&\overset{(d)}{\geq}\sum_{i=1}^{n}I(S_j,W_j,X^{i-1};\widetilde{X}_{j,i})\overset{(e)}{=}\sum_{i=1}^{n}I(U_{j,i};\widetilde{X}_{j,i})\label{eq:storageconv2}
\end{align}
where $(a)$ follows by (\ref{eq:storageconstJ2}), $(b)$ follows because $\widetilde{X}^n$ is independent of $S_j$ and from the encoding step, $(c)$ follows because the source and channel are memoryless, $(d)$ follows by applying the data-processing inequality to the Markov chain in (\ref{eq:MarkovchainXtildeiXiminus1}), and $(e)$ follows from the definition of $U_{j,i}$.

\emph{Proof for (\ref{eq:sumonsamesandwpart1Twoenroll})}: We have for the multi-enrollment GS model for $j=1,2$ that
\begin{align}
	&n(\widebar{R}_{\text{s},j}+\widebar{R}_{\text{w},j})\overset{(a)}{=}H(S_j,W_j)+I(S_j;W_j)+n\delta_n\nonumber\\
	&\overset{(b)}{\leq} \sum_{i=1}^n\big[H(S_j,W_j,X^{i-1})+\frac{\delta_n}{n}+\delta_n\big]\nonumber\\
	&\overset{(c)}{=} \sum_{i=1}^n\big[H(U_{j,i})+\frac{\delta_n}{n}+\delta_n\big]\label{eq: RsjRwjsumconversegs}
\end{align}
where $(a)$ follows by (\ref{eq:uniformityconstJ2}), $(b)$ follows by (\ref{eq:secrecyconstJ2}), and $(c)$ follows from the definition of $U_{j,i}$.

\emph{Proof for (\ref{eq:cssumonsamesandwpart1Twoenroll})}: 
Similarly, we have for the multi-enrollment CS model for $j=1,2$ that
\begin{align}
&n\widebar{R}_{\text{w},j}\leq \sum_{i=1}^nH(S_j,W_j,X^{i-1})\overset{(a)}{=} \sum_{i=1}^nH(U_{j,i})\label{eq:Rwjupperboundcs}
\end{align}
where $(a)$ follows from the definition of $U_{j,i}$.

\emph{Proof for (\ref{eq:sumonsamesjandwjandjprimeTwoenroll})}: We obtain for the multi-enrollment GS model for $j=1,2$ and $j'$ as defined in (\ref{jprimedef}) that 
\begin{align}
	&n(\widebar{R}_{\text{s},j}+\widebar{R}_{\text{w},j}+\widebar{R}_{\text{w},j'})\nonumber\\
	&\overset{(a)}{=} H(S_j,W_j,W_{j'})+I(S_j;W_j,W_{j'})+I(W_j;W_{j'})+n\delta_n\nonumber\\
	&\overset{(b)}{\leq}\sum_{i=1}^n \Big[H(S_j,W_j,W_{j'},S_{j'},X^{i-1}) + \frac{2\delta_n}{n}+\delta_n\Big]\label{eq:whereweneedindhelperdata}\\
	&\overset{(c)}{=}\sum_{i=1}^n \Big[H(U_{j,i},U_{j',i}) + \frac{2\delta_n}{n}+\delta_n\Big]\label{eq:highestsumrateconverse}
\end{align}
where $(a)$ follows by (\ref{eq:uniformityconstJ2}), $(b)$ follows by (\ref{eq:secrecyconstJ2}) and (\ref{eq:storageindependence}), and $(c)$ follows from the definitions of $U_{j,i}$ and $U_{j',i}$.

\emph{Proof for (\ref{eq:cssumonsamesjandwjandjprimeTwoenroll})}: We have for the multi-enrollment CS model for $j=1,2$ and $j'$ as defined in (\ref{jprimedef}) that 
\begin{align}
	&n(\widebar{R}_{\text{w},j}\!+\!\widebar{R}_{\text{w},j'})\nonumber\\
	&\leq\sum_{i=1}^nH(W_j,W_{j'},S_j,S_{j'},X^{i-1})+I(W_j;W_{j'})+n\widebar{R}_{\text{s},j'}\nonumber\\
	&\overset{(a)}{\leq} \sum_{i=1}^n\Big[H(W_j,W_{j'},S_j,S_{j'},X^{i-1})+\frac{\delta_n}{n}+\widebar{R}_{\text{s},j'}\Big]\label{eq:whereweneedindhelperdata2}\\
	&\overset{(b)}{=} \sum_{i=1}^n\Big[H(U_{j,i},U_{j',i})+\frac{\delta_n}{n}+\widebar{R}_{\text{s},j'}\Big]\label{eq:highestsumrateconversecs}
\end{align}
where $(a)$ follows by (\ref{eq:storageindependence}) and $(b)$ follows from the definitions of $U_{j,i}$ and $U_{j',i}$.

\begin{remark}\label{rem:helperdataindependence}
	\normalfont (\ref{eq:whereweneedindhelperdata}) and (\ref{eq:whereweneedindhelperdata2}) are the only places we use the constraint in (\ref{eq:storageindependence}) and it does not seem straightforward to obtain the inequalities in (\ref{eq:whereweneedindhelperdata}) and (\ref{eq:whereweneedindhelperdata2}) without (\ref{eq:storageindependence}).
\end{remark}

Introduce a uniformly distributed time-sharing random variable $\displaystyle Q\!\sim\! \text{Unif}[1\!:\!n]$ independent of other random variables. Define $X\!=\!X_Q$, $\displaystyle \widetilde{X}_j\!=\!\widetilde{X}_{j,Q}$, $\displaystyle Y_j\!=\!Y_{j,Q}$, and $U_j\!=\!(U_{j,Q},\!Q)$ so that $\displaystyle U_j\!-\!\widetilde{X}_{j}\!-\!X\!-\!Y_j$ form a Markov chain for $j=1,2$. The outer bound for the GS model follows by using the introduced random variables in (\ref{eq:secretkeyconv1}), (\ref{eq:convprivacylowerbound}), (\ref{eq:converseprivleakupperbound}), (\ref{eq:storageconv1}), (\ref{eq: RsjRwjsumconversegs}), and (\ref{eq:highestsumrateconverse}), and letting $\delta_n\rightarrow0$. Similarly, the outer bound for the CS model follows by using the introduced random variables in (\ref{eq:secretkeyconv1}), (\ref{eq:convprivacylowerbound}), (\ref{eq:converseprivleakupperbound}), (\ref{eq:storageconv2}), (\ref{eq:Rwjupperboundcs}), and (\ref{eq:highestsumrateconversecs}), and letting $\delta_n\rightarrow0$.

\section{Conclusion}\label{sec:conclusion}
We derived inner bounds for the multi-entity key-leakage-storage regions for GS and CS models with strong secrecy, a hidden identifier source, and correlated noise components at the encoder and decoder measurements that are modeled as BCs. The inner bounds are valid for any finite number of entities that use the same hidden source to agree on a secret key. We argued that the mutual key independence constraint we impose makes the proposed multi-entity key agreement problem a proper multi-user extension of the classic single-enrollment key agreement problem, unlike the multi-enrollment key agreement problem considered in the literature. A set of degraded and less-noisy BCs was shown to provide strong privacy without a need for a common randomness. We also established inner and outer bounds for the key-lekage-storage regions for a two-enrollment model with measurement channels that are valid for SRAM and RO PUFs. Inner and outer bounds were shown to differ only in the Markov chains imposed and they match if the storage and privacy-leakage rate constraints are removed. Two examples illustrated that depending on the constraints of the practical scenario, a single or multiple enrollments might perform better in terms of the secret-key vs. privacy-leakage rate ratio. In future work, we will find a set of symmetric probability distributions for which the strong helper data independence constraint in the two-enrollment model can be eliminated.  

\section*{Acknowledgment}
O. G\"unl\"u thanks Rafael F. Schaefer for fruitful discussions. 

\bibliographystyle{IEEEtran}
\bibliography{references}

\begin{IEEEbiography}[{\includegraphics[width=1in,height=1.25in,clip,keepaspectratio]{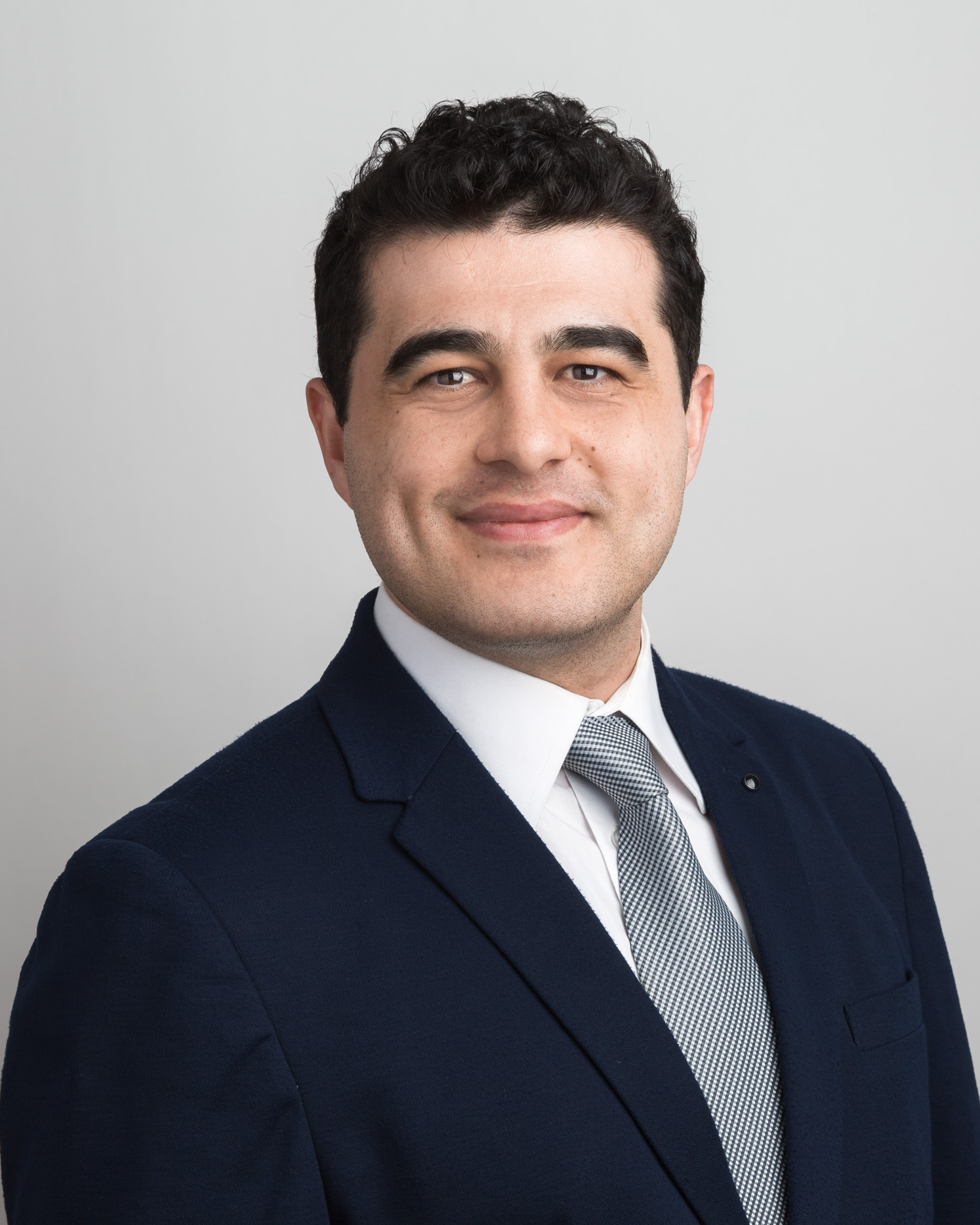}}]{Onur G\"unl\"u}
	(S'10--M'18) received the B.Sc. degree (with high distinction) in Electrical and Electronics Engineering from the Bilkent University, Turkey in 2011; M.Sc. (with high distinction) and Dr.-Ing. (Ph.D. equivalent) degrees in Communication Engineering both from the Technical University of Munich (TUM), Germany in October 2013 and November 2018, respectively. He was a Working Student in the Communication Systems division of Intel Mobile Communications (IMC) during November 2012 - March 2013. He worked as a Research and Teaching Assistant at TUM between February 2014 - May 2019. He was a Visiting Researcher at the Information and Communication Theory (ICT) Lab of TU Eindhoven, The Netherlands during February 2018 - March 2018. He has been a Research Associate and Dozent at TU Berlin, Germany since June 2019 and a Brain City Berlin Ambassador since June 2020. His research interests include information theoretic privacy and security, coding theory, statistical signal processing for biometrics and physical unclonable functions (PUFs), federated learning (FL) with differential privacy (DP) guarantees, and doubly-exponential secure identification. Among his publications is the recent book \emph{Key Agreement with Physical Unclonable Functions and Biometric Identifiers} (Dr. Hut Verlag, 2019). He is currently a Guest Editor of the \textsc{IEEE Journal on Selected Areas in Information theory} and is a Reviewer Board Member of the \textsc{MDPI Entropy}, \textsc{Computers}, and \textsc{Information} journals.
\end{IEEEbiography}

\end{document}